\documentclass[preprint,authoryear,5p]{elsarticle}
\usepackage[T1]{fontenc}
\usepackage{graphicx}
\usepackage{xcolor}
\usepackage{listings}
\usepackage{amsfonts, amssymb, amsmath}
\usepackage{cleveref}
\usepackage{caption}
\usepackage{subcaption}

\definecolor{draft}{RGB}{0, 0, 0}
\newcommand*{\Comment}[1]{\hfill\makebox[7.50cm][l]{#1}}%
\newtheorem{definition}{Definition}

\makeatletter
\def\@opargbegintheorem#1#2#3{\trivlist
   \item[]{\bfseries #1\ #2\ (#3)} \itshape}

\renewcommand\paragraph{\@startsection{paragraph}{5}{\z@}%
  {3.25ex \@plus1ex \@minus.2ex}%
  {-1em}%
  {\normalfont\normalsize\bfseries}}
\makeatother

\journal{}

\begin{document}


\title{NetGAP: A graph grammar approach for concept design of networked platforms with extra-functional requirements}
%
%

\author[1]{Rodrigo Saar de Moraes\corref{cor1} }
\ead{rodrigo.moraes@liu.se}
\author[1]{Simin Nadjm-Tehrani }
\ead{simin.nadjm-tehrani@liu.se}

\cortext[cor1]{Corresponding author}

\affiliation[1]{organization={Department of Computer and Information Science, Link\"{o}ping University},
            city={Link\"{o}ping},
            country={Sweden}}

\date{}


\begin{abstract}
During the concept design of complex networked systems, concept developers have to ensure that the choice of hardware modules and the topology of the target platform will provide adequate resources to support the needs of the application. For example, future-generation aerospace systems need to consider multiple requirements, with many trade-offs, foreseeing rapid technological change and a long period for realization and service. For that purpose, we introduce NetGAP, an automated 3-phase approach to synthesize network topologies and support the exploration and concept design of networked systems with multiple requirements including dependability, security, and performance. NetGAP represents the possible interconnections between hardware modules using a graph grammar and uses a Monte Carlo Tree Search optimization to generate candidate topologies from the grammar while aiming to satisfy the requirements. We apply the proposed approach to a synthetic version of a realistic avionics application use case. \textcolor{draft}{It includes 99 processes and 660 messages. The experiment shows the merits of the solution to support the early-stage exploration of alternative candidate topologies. The method vividly characterizes the topology-related trade-offs between requirements stemming from security, fault tolerance, timeliness, and the "cost" of adding new modules or links. We also create a scaled-up version of the problem (267 processes, 1887 messages) to illustrate scalability.} Finally, we discuss the flexibility of using the approach when changes in the application and its requirements occur.
\end{abstract}



\maketitle

\section{Introduction}
\par  Designing complex cyber-physical systems ~\citep{KhMc14} at the concept development stage can be a challenging task since there are a lot of uncertainties about the system at this stage. In Integrated Modular Avionics (IMA)  systems ~\citep{Watk06}, with a life-cycle over several decades and potential functionality changes, we need to estimate resource needs at the early stage while being flexible enough to absorb future modifications. \textcolor{draft}{ Considering an envisaged set of functionalities, it is crucial to account not only for the tasks they perform but also for extra-functional requirements \citep{Fair10}, such as performance, reliability, security, and cost-effectiveness. In practice, this involves finding a candidate computation and communication platform configuration that offers sufficient resources to support the intended application. This assessment includes checking if the selected hardware modules and their interconnections ensure ample computing and networking resources, while also allowing for the necessary safety and security measures within the topology to meet the specified requirements.}

\par Due to the complexity of the problem, several techniques have been proposed to tackle the problem of searching for adequate topologies in networked systems. Some works, such as \citet{GKDR06},  \citet{ZhZh07} and \citet{FBK+20}, use heuristic algorithms to generate industrial network topologies considering real-time constraints. Other works resort to more exact approaches, such as linear optimization or constraint satisfaction problems \citep{AnRT14}. An alternative approach is to use hybrid techniques, such as \citet{HeWG20}, which use a control-flow-based approach to generate architectures compliant with AFDX specifications and real-time performance requirements. 

\par Most of these approaches, however, focus on topologies of limited size, fixed structure, and little flexibility in the choices of hardware modules. Real-world systems are characterized by topologies configured with several types of hardware modules (e.g. heterogeneous processing modules, communication interfaces, remote data aggregators, and network switches) and an array of different connection arrangements. Therefore, there exists a need for an exploration technique that can capture the (distributed) topology generation problem generically, and navigate in this specific search space. Note that, for the remainder of the paper, we use the terms \textit{topology} and \textit{platform configuration} (or simply \textit{configuration}) interchangeably.

\par In this paper, we introduce NetGAP\footnote{ The name NetGAP blends elements of the words "Network", “Grammar" and "Approach"}, an automated approach to synthesize network topologies and support the exploration and concept design of IMA systems at the concept level. We represent each candidate topology as a graph generated through the use of a graph grammar 
to express possible arrangements of interconnections among hardware modules. Each candidate topology is derived by applying a sequence of grammar rules which, if applied in that order, generate the configuration in question.
Without the loss of generality, and to suit relevant IMA systems, we exemplify a graph grammar that is tailor-made to the types of topologies of interest for these systems. Using only a small set of rules, our grammar can describe thousands of possible alternative configurations composed of hardware modules of different types and their respective connection arrangements. 

\par Guided by a set of requirements imposed on both the application and topology, we steer the search through the explored space and decide on a sequence of rule applications to synthesize a topology that supports the envisaged application. In this paper,  we use a Monte-Carlo tree search technique \citep{CSB+06} to direct the search towards topologies that are likely to adhere to high-level requirements from the application. These requirements impose constraints on the platform modules, the mapping of application processes to them, and the chosen topology. More specifically, we consider aspects such as redundancy, network segmentation, computation and networking resource adequacy, estimated latency, and cost, while generating configurations that are suitable for the targeted avionics functionalities. 
\par We evaluate the proposed approach through the application to a realistic use case in the avionics domain, and discuss the scalability and expressiveness of NetGAP. Furthermore, we show how this approach can be used to provide insights on the merits of alternative platform configurations and support decision-making when conflicting requirements require a trade-off when selecting among different platform attributes or topologies.
\par In summary, this paper presents the following key contributions:

\begin{itemize}
    \item A recursive graph grammar to represent generic networked system topologies that enables the representation of a wide range of candidate concept designs.
    \item A hybrid approach leveraging Monte-Carlo-based methods and genetic algorithms to generate and optimize topologies given the requirements of an envisaged application.
\end{itemize}

\par The rest of the paper is organized as follows. \Cref{sec:related} gives an overview of related works dealing with similar problems. \Cref{sec:preliminaries} provides background on the techniques and concepts used in this paper. \Cref{sec:problemSolution} provides an overview of the problem and an overview of how NetGAP addresses it. \Cref{sec:grammar} provides a formal view of graph grammars and their implementation in this paper. \Cref{sec:platformSO} provides information on the optimization part of the problem and how the genetic and tree search algorithms are implemented.  \Cref{sec:evaluation} shows the application of the proposed methodology to a use-case and discusses issues such as the scalability and expressiveness of the approach. Finally, \Cref{sec:conclusion} concludes the paper and presents the future works.

\section{Related works \label{sec:related}}

\par For decades, the topological optimization of networks has been an important problem across many fields such as telecommunications \citep{ArBM15,HaMR08}, electricity distribution, gas pipelines, integrated circuit design \citep{SrCK05, DuKh09,LWG+18} and, recently, machine learning \citep{StMi02,ASAT15}. With the advent of complex distributed computing architectures within cars, planes, and modern data centers, the problem has seen new incarnations applied to computer networks. The ever-increasing complexity of modern networked systems, the expansion of the volume of data being exchanged, and the reliability and security requirements of contemporary and future applications have made the topology design of the local area and industrial communication networks a very complicated task. Current applications require all information to be available at all times and everywhere. As a result, legacy bus technologies and many common industrial protocols are quickly becoming obsolete, and switched networks have become the dominating local area network technology.

\par Traditionally, topology optimization of switched networked systems was the problem of choosing a set of links for a given set of nodes to optimize some performance measure such as economic cost, average message delay, network reliability, or a combination of them. Considering there are two kinds of nodes in LANs, network switches and end nodes, \citet{ElSi96} proposed the design of local area networks (LANs) can be divided into two sub-problems: allocating end nodes to switches and designing the interconnection between switches. The first problem tries to allocate end nodes that communicate frequently to the same switch (clustering problem), minimizing inter-switch communication and reducing network delays. The second problem deals with the choice of inter-switch communication (usually through the computation of spanning trees), guaranteeing that nodes connected to different switches can communicate and, if necessary, ensure reliability. This bipartite formulation has been extensively applied to industrial networks with hierarchical topology and is commonly known as the industrial Ethernet network partition problem (IENPP) \citep{GKDR06,ZhZh07}.

\par Due to the nature of the problem, the complexity of most algorithms proposed to solve the problem of topology design is exponentially dependent on the number of network nodes. Even though some works, such as \citet{AnRT14}, do propose the use of binary or mixed-integer linear programs to solve the problem, its exponential complexity makes it impossible to find a solution in reasonable time except for specific, often small, instances. Therefore, to design real-life-sized networks, several works propose to search for near-optimal solutions using (meta) heuristics instead. 

\par In general, heuristics can be iterative or constructive. Iterative heuristics usually start with an initial solution and repeatedly modify the candidate solution in each iteration. Meanwhile, constructive schemes build the solution part by part, in a piecemeal manner. While iterative schemes usually require longer computational time to achieve good results, constructive heuristics are usually fast but have to take decisions based on partial solutions, being very prone to being stuck in local minima. Iterative schemes, on the other hand, can avoid the trap of local minima by temporarily accepting suboptimal candidate solutions to improve upon. Unlike constructive algorithms, which only solve at the end of the design process, iterative algorithms produce many solutions with different qualities during their search. 

\par Among iterative heuristics, genetic algorithms (GA), briefly described in \Cref{sec:preliminaries}), are the preferred approach in the literature to solve the topology optimization problem, been applied extensively throughout the years. Earlier works, such as \citet{ElSi96, KrRT02, GKDR06,  ZhZh07, CcSP10} use different flavors of genetic algorithms to solve the problem in the context of hierarchical industrial Ethernet networks. Other works, such as \citet{ FBK+20}  and \citet{DeAS97} apply genetic algorithms to the context of distributed networks with mesh or mixed topology. Finally, \citet{ZhLZ10,ZhLW11} apply the concept of genetic algorithm to industrial multi-ring networks. One difficulty in designing an evolutionary algorithm for the problem is that genetic algorithms must be tailored to the characteristics of the problem at hand to perform well. Specifically, the choice of a suitable representation for feasible solutions is not trivial and makes the algorithm very sensitive to the problem it is being applied to. Furthermore, the choice of mutation, selection, and crossover operators, is crucial to the performance of the algorithm and depends on the chosen representation. In our context, this means that formulations of genetic algorithms built for different network topologies (i.e. hierarchical vs multi-ring) or different optimization objectives (cost, resilience, delay) can be significantly different and that no single formulation fits all cases. However, for solving a subpart of our problem that is not directly related to topology design, we still apply GA as described in section \Cref{sec:platformSO}. 

\par Apart from genetic algorithms, other iterative techniques can also be used. \citet{SDTS13} propose a special psychoclonal algorithm for the design of computer networks. Psycoclonal algorithms \citep{KPST06} are bio-inspired algorithms that leverage the idea of an artificial immune system to evolve a meta-heuristic to solve complex problems. Other common techniques such as simulated annealing (SA) have also been proposed in the context of designing industrial switched networks. \citet{PHBD95} proposes an SA formulation to solve the minimum spanning tree and node addition parts of the problem aiming to minimize cost and message delay on the proposed networks. Finally, \citet{YoSK02} proposes a hybrid fuzzy selective evolution technique to optimize the topology design of industrial networks for the same metrics. These formulations suffer from the same limitations as genetic algorithms and considerably restrict the architectures one can generate. In addition to being tailored to address a specific topology, these approaches do not allow for the addition or removal of nodes during the process. Instead, they try to find a spanning tree to connect existing nodes. Our approach, on the other hand, allows for the incremental addition, swap, and removal of nodes, which might be a necessary action to move towards better designs.


\par In the avionics domain, \citet{HeWG20} use a control-flow-based approach to generate architectures compliant with AFDX specifications and real-time performance requirements. The authors claim that the most common techniques for network topology generation, such as community discovery and topological grouping, cannot directly guarantee real-time performance. As an alternative, they propose an algorithm that starts by grouping end nodes (message generators and consumers) according to the measured centrality of the node (a measure related to the number of communication connections from/to the node and their delay tolerance)  and the number of proposed switches on the network. Each generated group is then connected to a switch and the inter-group delay tolerance guides the choice of which switches shall connect. \citet{AnRT14} on the other hand, proposes a binary formulation for a similar, but more complex problem. In addition to solving the switch interconnection problem by taking into account the monetary cost and reliability of switches and links, they also address platform deployment metrics such as the aircraft anatomy with installation locations and cable lengths and the total weight of the platform. None of these approaches, however, take into account other high-level extra-functional requirements in the concept design space.

\par Network topology design is also a commonly studied subject in the electronics industry, where a network-on-a-chip (NoC), needs to be optimized for power, area utilization, and performance while respecting very strict timeliness constraints. \citet{SrCK05} tackles the problem of designing custom application-aware NoCs with a three-phase optimization technique to create a performance-aware layout of the whole system-on-a-chip (SoC). \citet{DuKh09}  propose two algorithms to design custom NoCs to meet the communication requirements of an on-chip application while minimizing the utilization of network resources. Their multi-step approach starts by generating an initial topology and iteratively improving the solution until a good enough candidate platform is found. In a more recent study,  \citet{LWG+18} tackle a similar problem by clustering intra-chip cores between which the communication is heavy and thus should connect to the same switch or router, and proceed to generate a topology based on the number of cores in each cluster and a set of rules determining network growth. In the end, the authors apply a genetic algorithm to optimize the SoC floorplanning by minimizing power consumption and/or chip area.

\par In this work we combine the best of constructive and iterative techniques to create a topology-independent method that allows for the generation of mixed topologies and easy integration of different evaluation functions. Despite similarities, the main difference between the mentioned approaches and the one we present is the variety of requirements that can be represented and the large state space explored. While most approaches consider networks composed only of switches and application modules, our approach allows the inclusion of arbitrary user-defined types of network devices and modules and is scalable to a large number of such nodes as will be demonstrated in a real use case. Even though it is possible to expand some of the genetic algorithm formulations to consider different types of modules for small networks, their encoding usually poses a challenge for complex heterogeneous networks. 
Some approaches also consider that the number of network switches to include in the network is given beforehand. Our approach, through the use of the graph grammar approach, is naturally more flexible when it comes to the size of the networks and allows for the generation of multi-segmented networked platforms (i.e. multiple network segments connected by gateways or bridges) and mixed topologies (i.e. a single network containing a ring segment and a mesh segment).

\par Finally, the work in this paper approaches a different problem than the basic resource allocation analysis and trade-off studies when a \emph{given} topology is assumed (e.g. \citet{LEPB10,SmBN21,SmNt21}). The latter two works stop at the analysis of a given allocation candidate (subproblem 1 which will be described in \Cref{sec:problemSolution}) with the induced communication resource adequacy.
\section{Preliminaries \label{sec:preliminaries}}
\par In this section we present general definitions and background on the techniques and concepts used throughout the paper.

\subsection{Terminology}
\par It is not uncommon for different research communities to attribute different meanings to the same term or to use a different vocabulary to describe similar concepts. For clarity, we present here the terminology we use in the remainder of the paper.

\begin{itemize}
    \item \textbf{Functionality}: describes the intended goals of the system and the tasks it is supposed to perform.

    \item \textbf{Application}: describes the software components that realize a functionality. Applications are composed of software \textbf{processes} that exchange messages and work together to provide the expected behavior.
    
    \item \textbf{Platform:} the set of conceivable hardware and system software components on which applications can be hosted or deployed. \textbf{Platform modules} are thus the variety of hardware or low-level system software that support the execution of application processes. \textbf{Hardware modules} provide \textbf{resources} (e.g. communication bandwidth, memory, and computation capacity) used by the application processes in their computation or communication.
    
    \item \textbf{Platform Configuration} or \textbf{Topology}: the arrangement of a selected set of hardware modules according to a certain interconnection pattern. 
    
\end{itemize}

  Each topology will represent a candidate system architecture at the concept stage.

\subsection{Graphs and Grammars \label{sec:grammarDef}}

\par The process we will present in \Cref{sec:generation} consists of using a set of well-defined rules to transform a graph $G$ into another graph $H$. We will propose a grammar that is suitable for our goal in this paper. Before that, this section provides definitions adapted from  \citet{ Andries1999} and \citet{delaParra2014} of the key components necessary to formalize this process.

\subsubsection{Graphs and Subgraphs:}
\par In this paper, we consider edge- and vertex-labeled directed graphs as the basic structures for representing topologies. Graphs are formalized as follows:


\begin{definition}[Graph  \label{def:graph}]
A directed labeled graph over $\Sigma$ and $\Gamma$, where $\Sigma$ and $\Gamma$ are finite alphabets of distinct vertex and edge labels, respectively, is a 4-tuple $G = (V, E, \allowbreak l_V, l_E)$ where:
\begin{itemize}
    \item $V$ is a finite set of vertices
    \item $E \subseteq V \times V$ is a finite set of directed edges, from a source vertex to a target vertex
    \item $l_V: V \rightarrow \Sigma$ assigns a label to a vertex,
    \item $l_E: E \rightarrow \Gamma$ assigns a label to a vertex.
\end{itemize}
\end{definition}


\par Often, our interest is not to modify entire graphs, but specific parts of a graph, which we will henceforth call subgraphs. 

\begin{definition}[Subgraphs]
A graph $L$ is a subgraph of $G$, denoted by $L \subseteq G$, if the vertex and edge sets of $L$ are subsets of the respective sets of $G$, and the source, target, and label mappings of $L$, when restricted to the subsets in $L$, are identical to the respective mappings of $G$. 
\end{definition}

\subsubsection{Production Rules and Graph Grammars:} 

\par Intuitively, each rule application transforms a graph by replacing a part of it, described by the left-hand side (LHS) of the rule, with another graph, which we obtain from the right-hand side (RHS) of the rule. Formally:

\begin{definition}[Production Rules \label{def:production}]
A production rule is a 6-tuple $p=(L,K,R,g,e,a)$ that allows the transformation of a graph $G$ into a  graph $H$, where:

\begin{itemize}
    \item $L$ is the subgraph(s) identified by the left-hand side of the rule,
    \item $R$ is the subgraph(s) identified by the right-hand side of the rule,
    \item $K \subseteq L$ is a sub-graph of $L$ called the interface graph,
    \item $g$ is the occurrence of $K$ in $R$, relating the interface graph with the right-hand side graph,
    \item $e$ is an embedding relation relating vertices of $L$ to vertices of $R$,
    \item $a$ are the application conditions under which a production will be applied.
\end{itemize}
\end{definition}

\par
\begin{figure*}
    \centering
    \includegraphics[width=\linewidth]{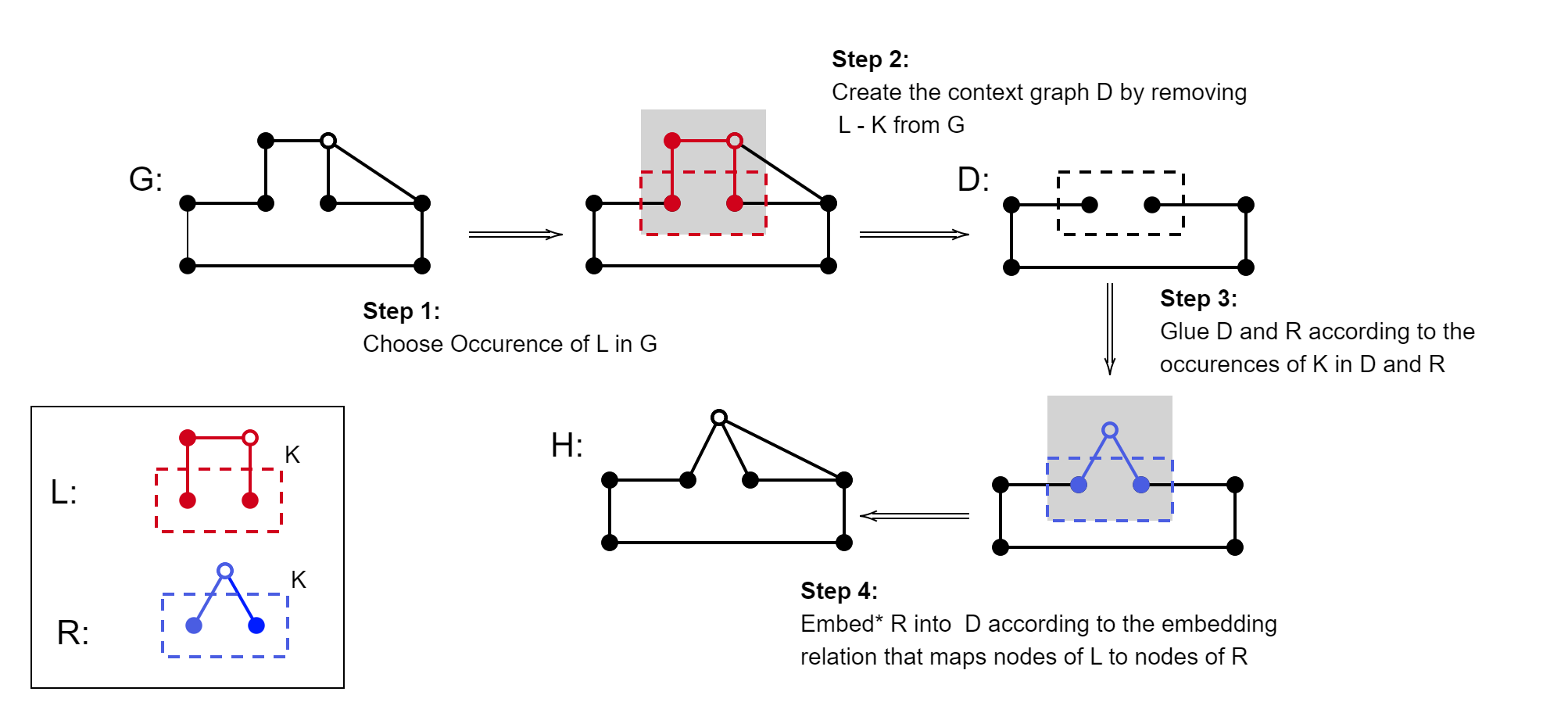}
    \caption{Illustration of the application of a production rule}
    \label{fig:production}
\end{figure*}

 The application of a production rule transforms a graph $G$ into a graph $H$ by creating a context graph $D =  (G-L)\cup K$ and gluing it to $R$ through $K$ ($H$ is essentially the disjoint union of graphs $D$ and $R$). \Cref{fig:production} illustrates this process to the reader (adapted from Fig 2. from \citet{Andries1999}). In the example of the figure, the embedding relation $e$ maps the unfilled vertex in $L$ to the unfilled vertex in $R$, thus the edges adjacent to the unfilled vertex in $G$, which are not part of $R$, are preserved in $H$.

\begin{definition}[Graph Grammar \label{def:graphgrammar}]
A graph grammar is a quadruple $GG = (T,S,P,A)$ where:
\begin{itemize}
    \item $T$ is the type or class of graph (e.g., labeled, directed, undirected), that can be generated, modified, and recognized by applying a set $P$ of productions.
    \item $S$ is the start graph to which a set $P$ of productions will be initially applied.
    \item $P$ is a set of productions $ \{ p1,..., pn \} $.
    \item $A$ is a set of additional specifications extending the scope and nature of applying the set of productions $P$ to graphs of type $T$ (e.g., attributes, programmed rules, global application conditions, structural conditions).
\end{itemize}
\end{definition}

\par Due to their wide diversity of applications, ranging from pattern recognition, compiler construction, model transformation, and generative design, graph grammars have received particular attention during the last few years. \cite{AlSK18} present a survey in the field of graph grammars in which they study and classify grammars according to the number of manipulated data, the nature of data, and finally the kind of data (images, graphs, patterns, etc.) manipulated by grammars. The way the A-term and the additional specifications on the topology are specified in the context of NetGAP will be described using the running case study in \Cref{sec:evaluation}.

\subsection{Monte Carlo Tree Search}
\par The idea behind using Monte Carlo-based methods for tree exploration was first introduced by \citet{CSB+06} for the evaluation of action values and game states applied to the game of Go. Since then, the technique has been generalized by \citet{CBSS08} as a unified artificial intelligence framework and applied to several other fields in which an actor has to choose an action based on the current state of a system \citep{BPW+12}. 
\par Monte-Carlo Tree Search (MCTS) is a best-first search technique that uses stochastic simulations to approximate the \textit{reward} of an action given the current state of a system (or game).  The framework is based on the idea of simulating future states of the system by applying a sequence of random actions from the current state until a termination condition is met. While a single random simulation does not offer much information about the quality of an action, exploring a  multitude of random simulations can provide valuable information on what the best action might be. As such, the goal of the simulations is to unfold the set of states that are reachable through the applications of actions to the current state, and simultaneously calculate a \textit{reward}  so that different explored actions and their respective accumulated rewards can be compared. 

\par
\begin{figure*}[ht!]
\centering
\includegraphics[width=0.8\textwidth]{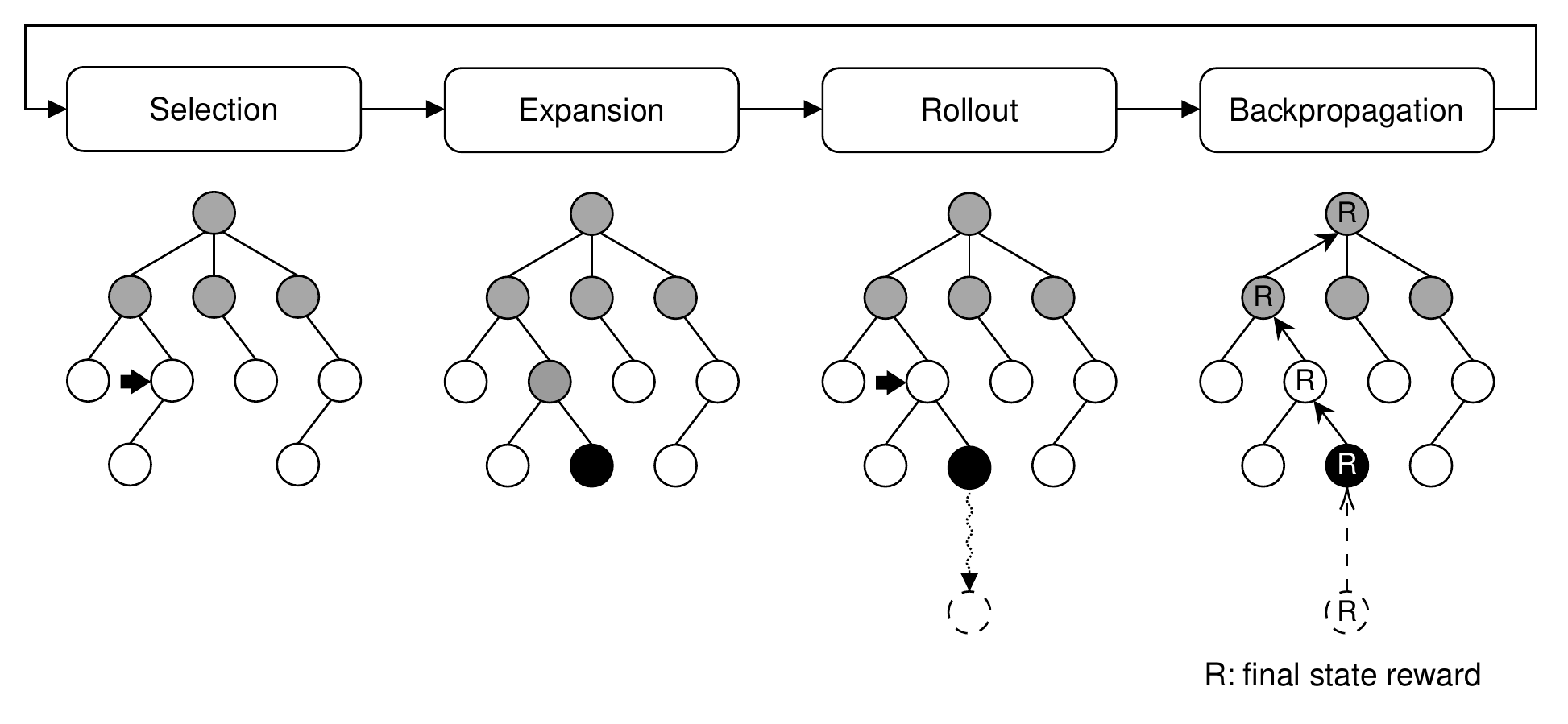}
\caption{ The MCTS algorithm. Grey shading indicates nodes that are fully expanded, white indicates nodes that are not fully expanded yet, and black shading indicates the current working node. } \label{fig:MCTS}
\end{figure*}

 MCTS, illustrated in \Cref{fig:MCTS}, operates by iteratively constructing and updating a tree of possible future system states using a combination of four strategic tasks: selection, expansion, rollout and backpropagation.  The \textbf{selection} operation, starts from the root and selects one existing leaf node $L$ to be expanded in a way that balances the search between exploitation and exploration. Exploitation means selecting the move that leads to the best results so far to optimize upon it. Exploration, in turn, means selecting less promising nodes that still have to be examined and may provide a good result in the future. The \textbf{expansion} operation adds new unexplored child nodes to $L$ based on the set of possible actions not tried from $L$. From the children of $L$, a node $C$ is chosen as the starting point, and random actions are taken during the \textbf{rollout} operation until a termination condition is met. The final state of the system after the random exploration (effectively a descendant of $C$) is not added to the tree. Rather, the final state is evaluated, and the reward earned by reaching that state is \textbf{backpropagated} through each tree node that was traversed during the simulation. Finally, the best action out of the current state (i.e. at the root of the tree) corresponds to that of the child node of the root with the best accumulated reward after the simulations are run.

\par Despite multiple strategies existing \citep{BPW+12}, the most common algorithm used for selection is UCT (Upper Confidence Bounds for Trees) \citep{KoSz06}. UCT, which is the method of choice in this study, improves upon the well-known UCB (Upper Confidence Bounds) \citep{ACbP02} algorithm to calculate the upper confidence bound of the reward associated with visiting each node of a tree and uses it as a criterion for selection. 
\par For the expansion task, which decides whether a leaf node will be expanded, the simplest rule is to expand one node per simulated game. Other strategies, however, such as expanding all the children of a node when a node’s visit count equals a given value \citep{CWH+08} are also possible. For the rollout task, which selects actions until the end of a simulation, the simplest approach is to select random actions at every step. This approach, however, does not lead to the best performance. Instead, one can use heuristics or an adequate set of rules to guide the random simulations to maximize the reward of an action (i.e. not choosing moves that will immediately lead to losing pieces in the games of chess). 

\subsection{Genetic Alogrithms}

\par Genetic Algorithms (GAs) \citep{Holl92}  are a type of iterative bio-inspired evolutionary algorithm that uses the ideas of Darwinian evolution to solve complex optimization problems. The concept behind these methods is to evolve an initial population of candidate solutions until either one of the solutions in this pool or population is considered good enough or the maximum number of iterations has been reached.
\par  The idea behind genetic algorithms is the application of a survival of the fittest principle on a population of individuals representing potential solutions to the problem being solved. By this principle, only the fittest individuals can survive and mate, causing potential solutions to converge toward the optimal solution. As in real life, each individual in a population has a unique genome that defines its traits and dictates its ability to solve the problem. Through a series of operations such as selection, crossover, and mutation, the genetic code of the population is evolved for a certain number of generations, or iterations, until a stopping condition is met.
\par The process of evolving a solution starts with the selection of a representation of the problem in terms of a genome, often called problem encoding. A suitable encoding should make sure that each candidate solution has its own representation while keeping important characteristics of the problem (such as the order of the genes). From that point, an initial population of candidate solutions is created at random or through heuristics, depending on the problem and domain-specific constraints. Each candidate of the initial population is analyzed, its fitness, or how well the candidate solves the problem, is calculated, and the population is sorted according to fitness values.  
\par To generate each individual in the next generation, two parents among the candidates are selected to originate offspring in a process we call crossover. During crossover, the genetic material of the parents is combined into new genomes to create one or more new individuals. Next, the mutation occurs. During mutation, some of the newly generated individuals are selected to suffer mutations in their genetic code, adding new unique traits to the genetic pool. The selection, crossover, and mutation process is independently repeated until a new population of the desired size is created. 
\par Once a new population is created, the algorithm applies the survival of the fittest principle again on this new generation and the crossover and mutation process repeats itself. This process is iterated over and over until either a stop condition is found or the maximum number of generations is exceeded. Ultimately, the best solution to the problem is represented by the genome of the fittest individual found during the evolutionary process.
\par One of the most important characteristics of heuristic algorithms is their ability to avoid poor local optima and converge towards near-optimal solutions. In genetic algorithms, the balance between exploration and exploitation is achieved through the choice of crossover, parent selection, and mutation operators. This choice, however, is not trivial and is highly dependent on the problem and the choice of encoding.

\section{Problem Statement and Solution Overview \label{sec:problemSolution}}
Given an envisaged set of functionalities, our task is to support the exploration of candidate platform configurations at the concept level.

\subsection{Problem Definition}

\par At the concept level, we are not concerned with finding optimal designs, instead, our goal is to explore the search space and identify which configurations are most likely to support the requirements of the application and which are their common characteristics. In particular, our interest lies in the search for topologies that ensure the adequacy of computing and networking resources and that provide safety and security mechanisms that meet the needs of the applications. The analysis at this stage is likely to uncover some conflicts or trade-offs among the system-level requirements when application processes are mapped to certain modules, and connected in certain patterns.

\textcolor{draft}{Furthermore, at the concept stage, we consider obtaining a rough estimation of resource adequacy and identifying which platform configurations are likely to result in feasible solutions is more important than obtaining detailed results such as scheduling plans for software processes or network messages. Detailed scheduling problems are then addressed at the detailed design stage when solutions to these non-trivial joint problems can be found using domain\linebreak-specific algorithms, see \citep{KRSU21}.  }

\par Following our previous work \citep{SmBN21}, a model on the software side, denoted by AM, describes the relevant computation and communication characteristics of a set of applications. Each application in the avionics application model (AM) is composed of periodic processes that communicate with each other through messages. The period of each process defines the frequency at which it will send its (output) messages. When deployed, processes are allocated to run on processing modules, which in this paper are referred to as hardware modules capable of computation. These may be, among others, system-on-a-chip boards, microprocessors, microcontrollers, video processors, and traditional processors. 

\par In addition, a topology is modeled as a directed labeled graph (see \Cref{def:graph} in \Cref{sec:grammarDef}), where the nodes represent hardware modules and edges represent communication links between these modules. Since a typical hardware module hosts application processes, messages exchanged between processes leave the source processing modules through the communication links and reach the destination modules through a direct connection between modules or through intermediary communication hardware modules (i.e. switches, routers, gateways, etc.). 

\par Therefore, the search for a candidate topology translates into the generation of a topology graph and a mapping of the application processes to the elements of the graph to meet the requirements imposed on the application and topology. Considering the complex nature of the problem, we divide it into three sub-problems: 

\begin{enumerate}
    \item The process allocation sub-problem (SP1): consists of finding a suitable allocation of software processes (with their associated communication needs) to processing modules to be consistent with the computation and communication capacities of the processing modules.  
    \item The topology generation sub-problem (SP2): consists of building the topology graph by deciding how many and which types of modules (processing modules, communication modules, etc.) should be present in the topology and how they should interconnect to respect the safety (fault tolerance), timeliness, and security related requirements.
    \item The mapping sub-problem (SP3): consists of finding a mapping between the solutions of the first two problems. This means establishing which processing modules in the solution SP1 (now hosting processes) correspond to which hardware module in the topology generated as a solution SP2.
\end{enumerate}

\subsection{NetGAP Overview \label{sec:solutionOverview}}

\par Given that all three sub-problems described in the previous subsection are NP-hard, trying to solve all three of them simultaneously may not be feasible and results in significant computational complexity. Therefore, in NetGAP, we address the ultimate goal by solving the three sub-problems iteratively. 

\par
\begin{figure*}[thb!]
\centering
\includegraphics[width=\textwidth]{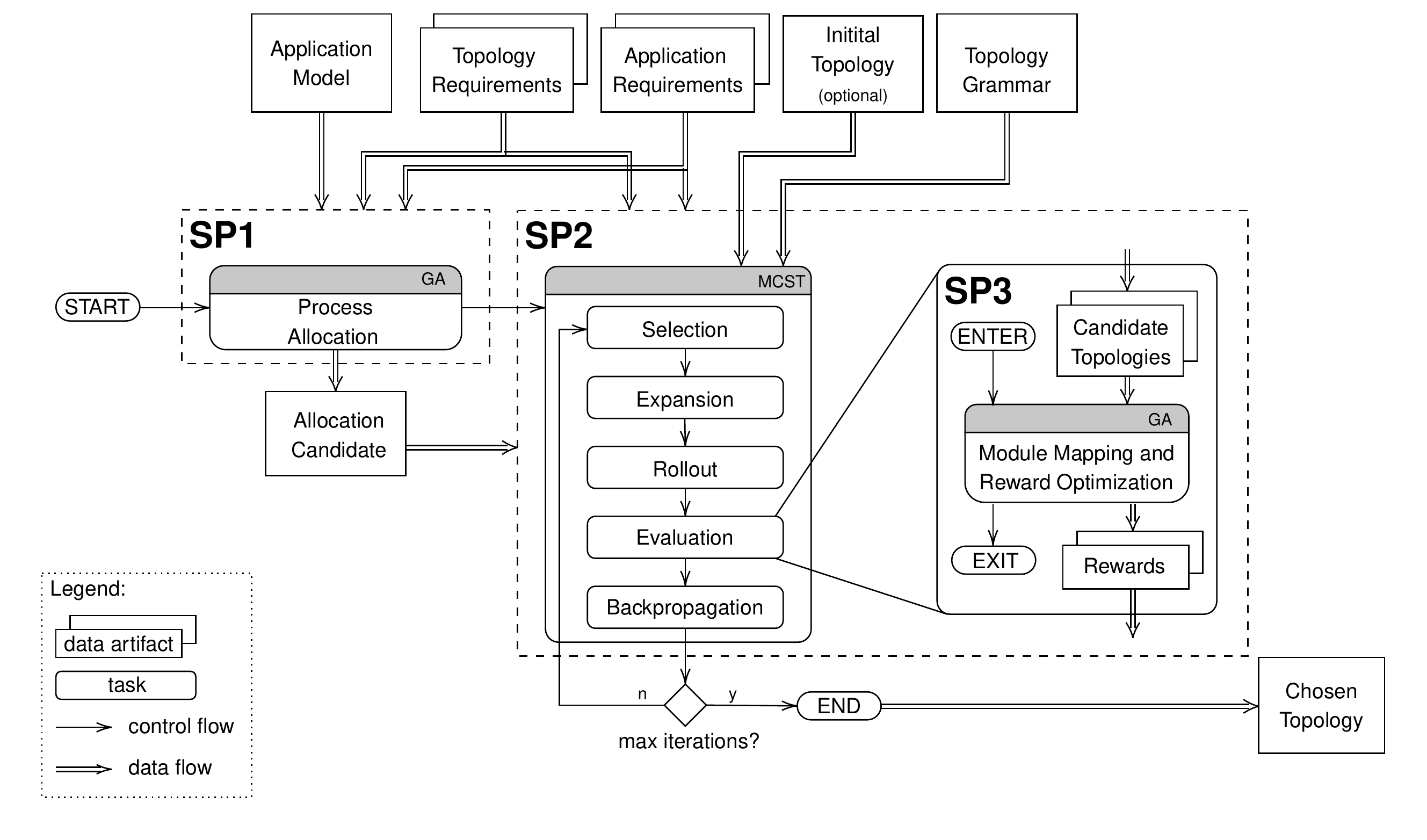}
\caption{NetGAP Overview.} \label{fig:solutionOverview} 
\end{figure*}

 \Cref{fig:solutionOverview} shows an overview of NetGAP and the artifacts/documents involved in the process. We also use the figure to highlight the stages in which the three sub-problems (identified by SP1, SP2, SP3) are situated and the algorithms used to address each of them (shaded nodes). The proposed approach takes four obligatory inputs: the topology grammar which describes possible interconnection patterns of the topology as general rules; the application model; a set of application requirements; and, lastly, a set of topology requirements  (e.g. max number of modules, cost, energy usage, etc.); and one optional input in the form of an initial topology to use as a starting point (if no initial topology is provided the empty topology is assumed.)

\par The NetGAP workflow starts with the process allocation stage (SP1), where a genetic algorithm is used to establish an optimized allocation of processes to processing modules, (henceforth called an \textit{allocation candidate}) and the required number of processing modules to host the application processes. Note that the allocation candidate also captures the inter-module communication patterns since from this point on we can easily pinpoint which modules messages travel from and to.
\par Using the output of SP1 to limit the search space, we solve the second and third problems iteratively using Monte-Carlo Tree Search. In SP2 we combine a topology grammar and the MCTS search to generate a sequence of candidate topologies based on the input from SP1, i.e. a set of (processing) hardware modules, and the required connections between them due to the application mapping from SP1. For each candidate topology, we map the application processes and their communication onto the topology (SP3) using a fast genetic algorithm and evaluate the merits  (henceforth called \textit{reward}) of this configuration in terms of the envisaged requirements. We feed the reward back into the MCTS loop and repeat the process for a predefined number of iterations (a configurable parameter to NetGAP) to generate the candidate topology considered as final. 
\par The process of generating a topology graph using the MCTS approach and the interaction between the steps will be discussed in depth in \Cref{sec:platformSO}. The next subsection provides an intuitive overview of the grammar-based graph generation process.

\subsection{Grammar-Based Topology Graph Generation \label{sec:generation}}

\par
\begin{figure*}[htb!]
\centering
\includegraphics[width=\textwidth]{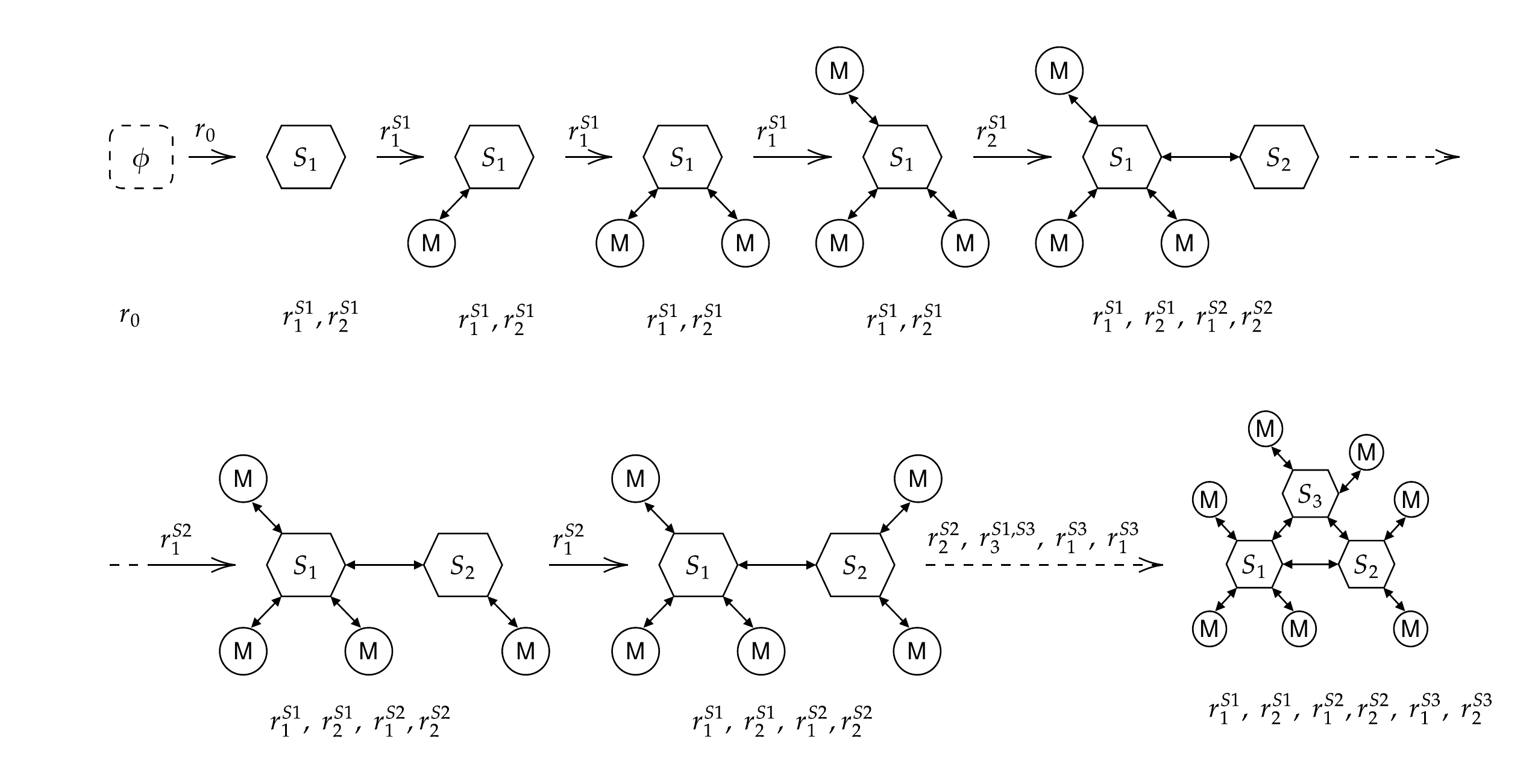}
\caption{Grammar-based generation of a topology using the grammar of \Cref{grammarlst1}. The list of rules below the diagrams indicates the set of possible rules that can be applied to the graph at each step of the process. Meanwhile, the labels above the arrows indicate the rule applied to proceed to the next step. Subscripts represent the rule number and superscripts represent the different nodes a rule can be applied to.} \label{grammarExample}
\end{figure*}

 In our approach we use a graph grammar to express possible interconnection patterns of platform modules. Each topology candidate is derived by applying a sequence of grammar rules which, if applied in that order, generate the platform configuration in question. An example of such a grammar is given in \Cref{grammarlst1}, \textcolor{black}{where the semantic interpretation of the rules is represented as a comment starting by \#}.  \Cref{grammarExample} illustrates the iterative process of generating a network topology using this example grammar.

\begin{lstlisting}[caption={A simple grammar example} ,mathescape=true,label={grammarlst1}, escapechar=\& ,basicstyle=\scriptsize,breaklines=true]
r0: $\phi \Rightarrow S$ &\Comment{\# Adds a switch node to an empty graph}&
r1: $S \Rightarrow$ S $ \leftrightarrow M$ &\Comment{\# Adds a module $M$ to the graph and connects it to switch $S$}&
r2: $S \Rightarrow S \leftrightarrow S'$ &\Comment{\# Adds switch $S'$ to the graph and connects it to switch $S$}&
r3: $S , S' \Rightarrow S' \leftrightarrow S$&\Comment{\# Connects two previously unconnected switches $S$ and $S'$ }&
\end{lstlisting}

\par
\begin{figure*}[hbt!]
\centering
\includegraphics[width=0.75\textwidth]{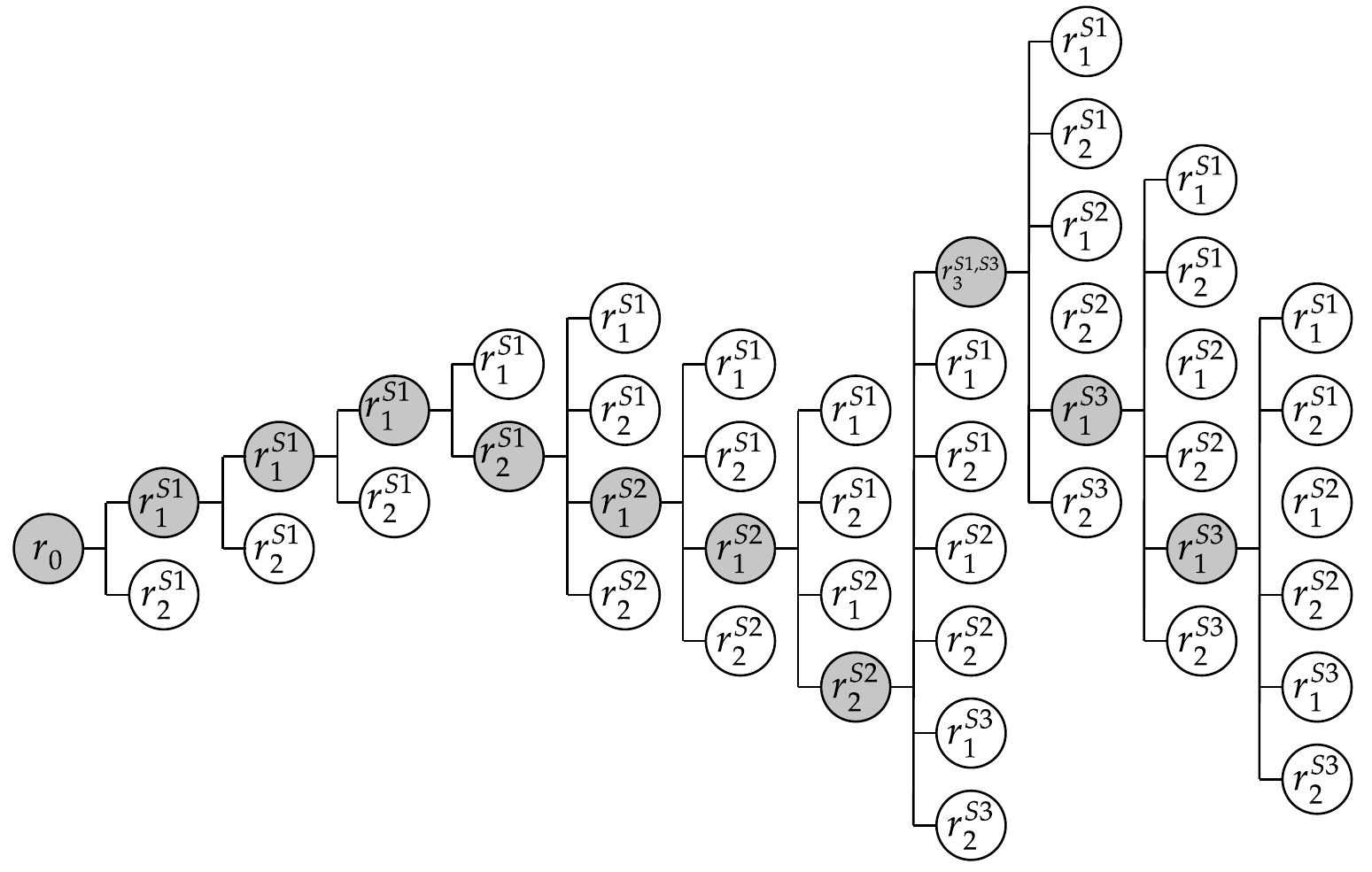}
\caption{A tree representing the possible rules available at each step of the topology generation process illustrated in \Cref{grammarExample}. Shaded tree nodes indicate the rules applied to proceed to the next step. Subscripts represent the rule number and superscripts represent the different nodes a rule can be applied to.} \label{grammarTree}
\end{figure*}

 As illustrated in \Cref{grammarExample}, at each step of the process there might exist multiple applicable rules depending on the structure and components of the graph at that moment. Choosing which rule to apply at each step is a non-trivial problem that tends to become intractable as the size of the graph and the complexity of the grammar increase. \Cref{grammarTree} shows a graphical representation of the possible rules applied from \Cref{grammarlst1} at each step, where subscripts represent the rule number and superscripts represent the different nodes a rule can be applied to ($r_1^{S2}$ represents the application of rule 1 to node S2). The shaded nodes represent the rules taken to generate the topology of \Cref{grammarExample} and the white nodes represent branches that were not expanded/explored.

\par To steer the search toward topologies that are likely to support the desired application, we need to evaluate the characteristics of the platform configuration at every step and analyze what the best rule to apply at that point is. The choice of a rule should reflect not only the current state of the graph but also how the application of said rule affects the partial and final candidate topologies generated from that point on. Hence we need to look ahead in the tree and steer the search toward configurations that are likely to respect the envisaged requirements. 

\section{A Grammar for Generic Networked Systems  \label{sec:grammar}}
\par In this section we define our notion of graph grammar for networked systems following the original definitions of generic graph grammars (Definition 4). We then go on to illustrate example applications of specific grammar rules, and the added syntax to give us the expressivity that we need.
\par Our network topology graph grammar is a tuple $\mathcal{N}=(G,S,P,A)$ where $G$ is always a labeled directed graph, $S$ is the starting graph (may be the null graph), and the sets of productions $P$ and additional specification $A$ are written according to the lexical rules in \Cref{lst:grammarDefinition} (note that the character '|' is used in the rules to distinguish between two alternative ways to rewrite a term).

\begin{lstlisting}[caption={Lexical rules of the network topology graph-grammar} ,label={lst:grammarDefinition}, basicstyle=\scriptsize, mathescape=true ]
<production> := <lhs> '$\Rightarrow$' <rhs>';' 
<rhs> := <list> 
<lhs> := <list> 
<list> :=  <struct> | <list>','<list>       
<struct> := <node> | <edge>
<edge> :=  <struct>'->'<struct> | <struct>'<->'<struct> 
<node> :=  <typelabel> | <typelabel> <interval>
<interval> := '['<number>']' | '['<number>','<number>']'
<typelabel> := '[a-zA-Z]+'
<number> := '[0-9]+'
\end{lstlisting}

\par In our formulation, a production rule is composed of a left-hand side (LHS) and right-hand side (RHS) separated by a $\Rightarrow$ symbol. 
Both the left and right-hand sides of a production formulation are composed of lists of subgraph structures separated by "," which, in turn, can be either single nodes or edge structures in the form A$\rightarrow$B, meaning an edge between A and B (or a sequence of node-edge structures of the type  A$\rightarrow$B$\rightarrow$...$\rightarrow$C). \textcolor{draft}{Structures of the type A$\leftrightarrow$B indicate that A and B are connected by edges in both directions}. Nodes are identified by their type labels which indicate the type of node and differentiate between two nodes of the same type if necessary. \Cref{lst:grammarExamples} shows some example rules that can be created using this formulation and their semantic interpretations expressed as comments (starting with \#).

\par When applying a production rule to a graph $G$, all structures in the LHS are matched to all corresponding structures in $G$, generating a list of all possible interface graphs (see \Cref{def:production}) to which the production can be applied. In general, elements in the LHS that are not present in the RHS are removed from $G$, while all elements in the RHS that are not in the LHS are added to the $G$. Therefore, productions such as A,B $\Rightarrow$ A will remove a node of type B from $G$ (provided two nodes of type A and B, respectively, are present anywhere in $G$). Meanwhile, production such as A,B $\Rightarrow$ A,B,C will add a node of type C to the graph (wherever two nodes of type A and B, respectively, are present anywhere in $G$). The only exceptions are constructions of the type A $\Rightarrow$ B which will transform a node of type A into a node of type B.

\begin{figure*}[ht!]
\centering
\includegraphics[width=0.8\textwidth]{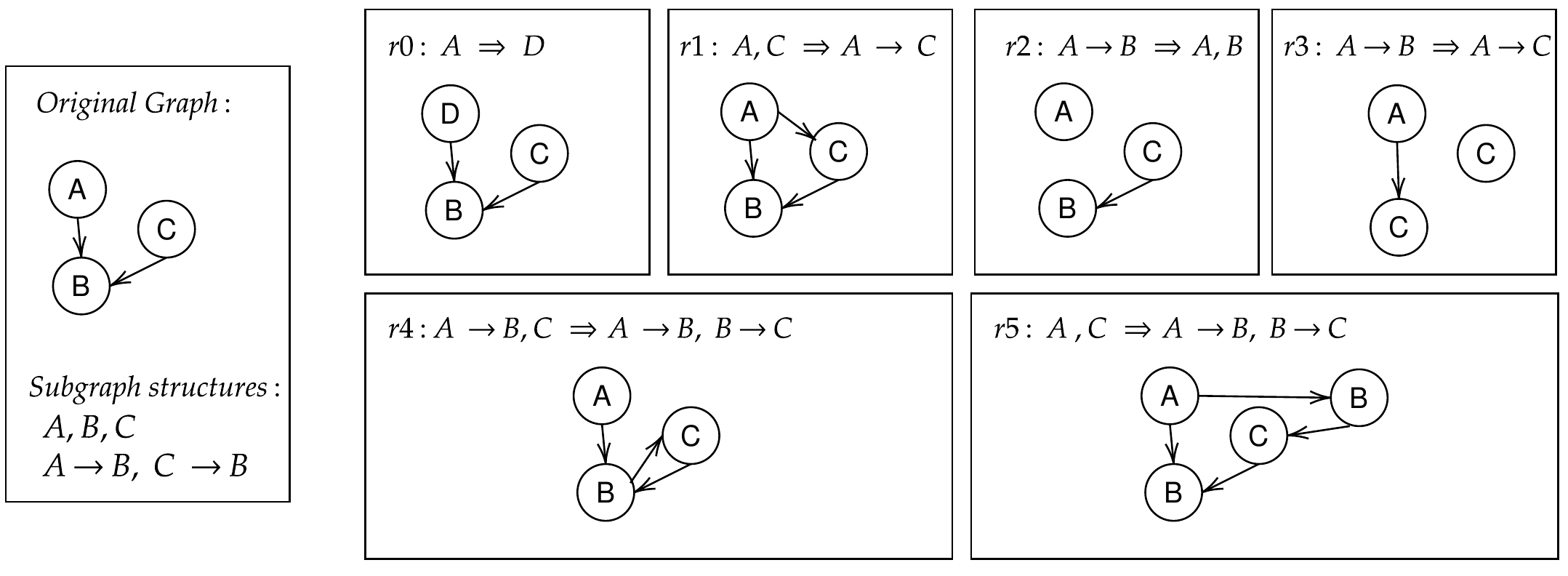}
\caption{Example of the application of rules r0 to r5 (one at a time) of \Cref{lst:grammarExamples} to an original graph seen on the left. } 
\label{fig:grammarApplication}
\end{figure*}

\Cref{fig:grammarApplication} shows six different applications of example rules r0 to r5 of \Cref{lst:grammarExamples} (one at a time) to the same graph with three nodes A, B, C and two edges $A \rightarrow B, C \rightarrow B$. Note that the rule examples in \Cref{lst:grammarExamples} are crafted to completely cover the rule writing syntax given our definition in \Cref{lst:grammarDefinition}. The goal is to introduce the reader to the semantic interpretation of different (valid) rule types. Some rules can be seen as syntactic sugar when the other rules are present, but our goal here is to show how each syntactic component in \Cref{lst:grammarDefinition} can be interpreted to express some operation of interest.

\begin{lstlisting}[caption={Examples of grammar rules and their semantic interpretation when applied to a generic graph G. Each rule in this example illustrates one type of graph transformation independently from other ones. },mathescape=true, breaklines=true, escapechar=\& , label={lst:grammarExamples}, basicstyle=\scriptsize]

r0:$A \Rightarrow D; $  &\Comment{\# replaces node A by node D}&
r1:$A,C \Rightarrow A \rightarrow C; $&\Comment{\# adds A $\rightarrow$ C if A,C $\subseteq$ G}&
r2:$A \rightarrow B \Rightarrow A, B; $&\Comment{\# removes the edge A $ \rightarrow $ B}&
r3:$A \rightarrow B \Rightarrow A \rightarrow C;  $&\Comment{\# if A $ \rightarrow $ B $\subseteq$ G, removes B and C$\rightarrow $B, adds C and  A $ \rightarrow $  C }&
r4:$A \rightarrow B,C\Rightarrow A\rightarrow B, B\rightarrow C;$ &\Comment{\# connects B to C if A $ \rightarrow $ B $\subseteq$ G}&
r5:$A,C \Rightarrow A \rightarrow B,B \rightarrow C;$  &\Comment{\# adds B, A $ \rightarrow $ B and B $ \rightarrow $ C if A,C $\subseteq$ G}&
r6:$A_{[8-10]},C \Rightarrow A \rightarrow C; $&\Comment{\# adds A $\rightarrow$ C if A,C $\subseteq$ G and $deg$A $\in [8,10]$}&
r7:$C_1 \rightarrow C_2 \Rightarrow C_1 \rightarrow B \rightarrow C_2;$ &\Comment{\# adds a node B between $C_1$ and $C_2$}&
\end{lstlisting}

\par For the envisaged types of topologies in IMA systems, we also define additional application conditions (see \Cref{def:production}) of the production rules to  allow for the selection of nodes with a specific degree. As an example, the formulation $A_{[8-10]}$ in r6 of  \Cref{lst:grammarExamples} defines the selection of a node of type A with a degree in the interval $[8,10]$, meaning a node with 8 to 10 connecting edges in any direction. This extension is especially useful for the description of a networked system in which the number of current edges to a switch or gateway node defines whether or not new connections can be allowed.

\par \textcolor{black}{In addition, we define an additional specification to our graph grammar (to exemplify structural conditions from \Cref{def:graphgrammar}, part A)  which  allows for the distinction of different nodes of the same type (i.e. node labels $C_1$, and $C_2$ in r7 both denote nodes of type C)}

\par
\begin{figure}
     \centering
     \begin{subfigure}[b]{0.3\textwidth}
         \centering
         \includegraphics[width=\textwidth]{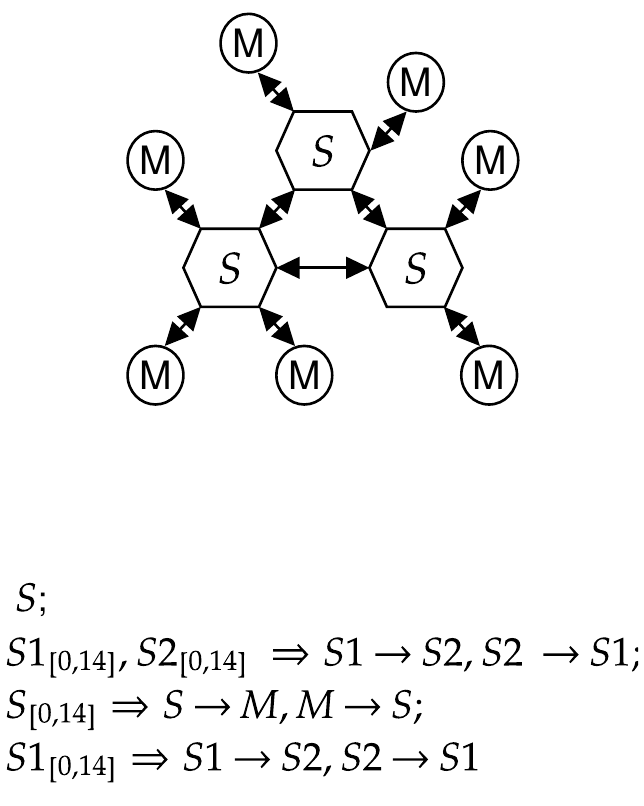}
         \caption{Mesh topology}
         \label{fig:8pmesh}
     \end{subfigure}
     \hfill
     \begin{subfigure}[b]{0.3\textwidth}
         \centering
         \includegraphics[width=\textwidth]{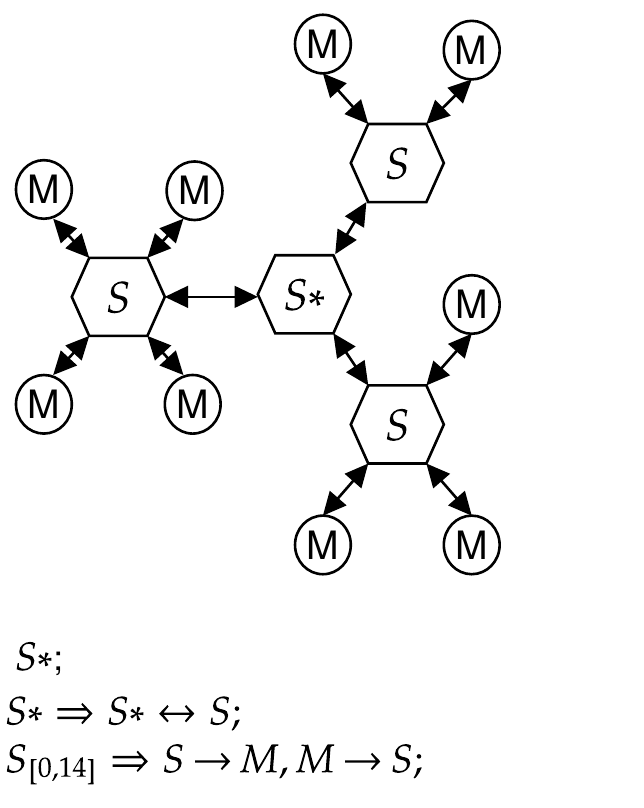}
         \caption{Extended star topology}
         \label{fig:8pestar}
     \end{subfigure}
     \hfill
     \begin{subfigure}[b]{0.3\textwidth}
         \centering
         \includegraphics[width=\textwidth]{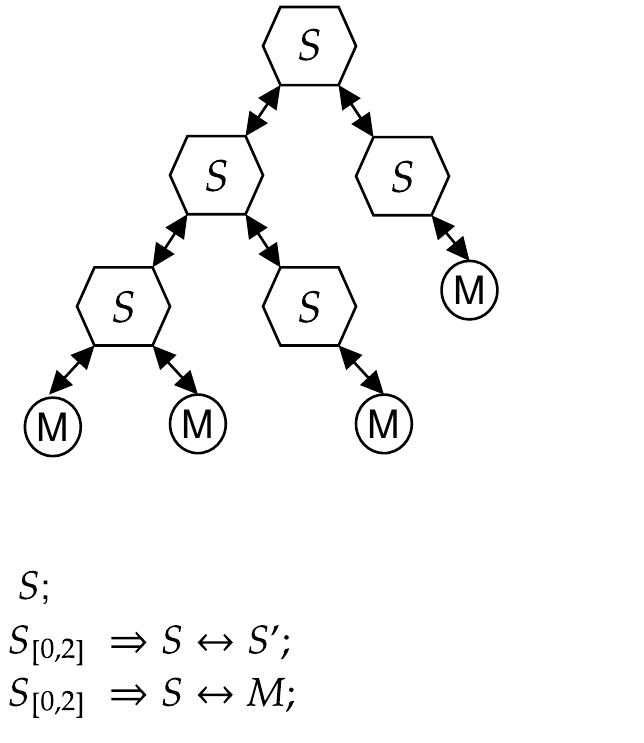}
         \caption{Tree topology}
         \label{fig:2ptree}
     \end{subfigure}
        \caption{Three network topologies and their respective grammar representations}
        \label{fig:grammarsAndNets}
\end{figure}

 Finally \Cref{fig:grammarsAndNets} depicts 3 different topologies (mesh, extended-star and tree) and examples of the respective grammar rules used to represent the specific characteristics of each.

\section{Topology Search and Optimization \label{sec:platformSO}}

\par In this section we discuss in detail our approach to solving each of the subproblems from \Cref{sec:problemSolution} (\Cref{fig:solutionOverview}).

\subsection{SP1: Process Allocation Problem \label{sec:processAllocation}}

\par Let $P=\{p_1,...,p_m\}$ be a finite set of processes and $N=\{n_1,...,n_q\}$ be a finite set of processing modules. Our first constraint is that we want to allocate each process $p_i \in P$ to one and only one module $n_j \in N$. Assuming $ x_{ij}$ is a binary variable that represents the allocation of process $i$ to module $j$, we can formulate this constraint as:
\begin{equation}
    \sum_{j} x_{ij} = 1      \quad\quad for\;all\; i \in P \\     
\end{equation}

\par Note that, processes can only be allocated to modules that exist in the platform configuration, therefore, considering $ y_j$ is a binary variable that represents the presence of module  $j$ in the system, we have our second constraint:

\begin{equation}
    \sum_{i \in P} x_{ij} \leq M y_j        \quad\quad for\;all\; j \in N \\
    \label{eq:moduleexistence}
\end{equation}

where $M$ should be at least larger or equal to the number of processes.

\par  The sum of the computation resources (represented as $r_i$ for process $p_i$) used by the processes assigned to a module $n_j$  must not exceed the computation resource capacity $w_j$ of that module, therefore:

\begin{equation}
    \sum_i x_{ij}r_i \leq w_j        \quad\quad for\;all\; j \in N 
\end{equation}

\par Consider module $n_j$ has incoming and outgoing bandwidth capacities $v^{in}_j$ and $v^{out}_j$, respectively. If two processes have been allocated to different modules, the communication between the processes will make use of the incoming and outgoing bandwidth capacities of the modules they are allocated to. On the other hand, if two processes $p_i$ and $p_k$ have been allocated to the same module $n_j$, the messages exchanged between them should not count towards the bandwith usage on the interfaces of the respective modules. Let $s_{ik}$ be the bandwidth requirement of the communication between processes $p_i$ and $p_k$, in that order. Assuming $z_{ikj}$ is a binary variable that represents the allocation of process $p_i$ to the same module $n_j$ as process $p_k$, the following constraints ensure that both the incoming and outgoing communication capacities are respected for each module:

\begin{equation} \sum_i \sum_k s_{ik}(x_{ij}-z_{ikj}) \leq  v^{out}_j       \quad\quad for\;all\; j \in N \end{equation}   
                    
\begin{equation} \sum_k \sum_i s_{ik}(x_{kj}-z_{ikj}) \leq  v^{in}_j      \quad\quad for\;all\; j \in N \end{equation}  

\par The following constraints ensure that $z_{ikj}$ equals to 1 if and only if both processes $p_i$ and $p_k$ have been allocated to the same module $n_j$.
\begin{equation} z_{ikj}\leq x_{ij}          \quad\quad for\;all\; j \in N, \;i,k \in P\end{equation}
\begin{equation} z_{ikj}\leq x_{kj}          \quad\quad for\;all\; j \in N, \;i,k \in P \end{equation}
\begin{equation} z_{ikj}\geq x_{ij}+x_{kj} -1          \quad\quad for\;all\; j \in N, \;i,k \in P\end{equation}

\par The final constraint is the integrality constraint that restricts the decision variables to binary values.

\begin{equation}
     x_{ij},y_{j},z_{ikj} = 1 \; or \; 0
\end{equation}

\par Our primary objective is to minimize the total cost of the processing modules necessary to host our processes, subject to the constraints formulated in Equations (1) to (9). Associating a cost for $c_j $ to each module, the main objective function is:

\begin{equation}
\label{eq:primObj}
     minimize \quad \sum_{j} y_jc_j \\
\end{equation}

\par If desired, secondary objectives can also be added. As an example, we could also try to minimize the computation and communication load on individual modules, balancing the computation and communication between processing modules. 
\par In the next subsection, we describe how we use genetic algorithms to solve the problem in a timely fashion. The choice for a heuristic over exact techniques, (e.g. a Mixed-Integer Linear Problem formulation) is due to genetic algorithms being more suitable to accommodate future changes to the objective function, even supporting non-linear or multiple/secondary objectives if desired.

\subsubsection{Genetic algorithm encoding of the allocation problem \label{sec:allocationProblem}}

\par Our genetic encoding of the problem sees the genome of a candidate solution decomposed into two separate parts, one representing the inclusion of modules and one representing the allocation of processes to modules. The first part of the genome is a list of booleans, each representing the inclusion (or not) of a respective module in the system. The second part of the genome is a list of integers, the position of each representing a process and the value of each representing the index of the module the process is allocated to in that solution (i.e the first integer on the list indicates which module $p_1$ is allocated to). 

\par When it is time to perform the mutation and crossover operations, each part of the genome is individually operated on. First, a simple two-point crossover is applied to the first part of the genome, then, the same operator is applied to the second part of the genome.  Given the independent nature of the two parts of the genome, the described crossover operations can generate an invalid offspring by allocating processes to modules that are not included in the working set. To account for that, the crossover operator for the second part of the genome has been modified as to generate only valid solutions. This is done by checking whether the allocation of every process is valid and correcting if necessary (either by inheriting that gene from a different parent or at random). After the crossover, a simple random mutation operation is used for both parts of the genome. Again, the operator for the second part of the genome has been modified (in the same fashion as the crossover) to generate only valid solutions.

\par To accelerate the process, the multi-objective optimization is performed in two steps. First, the entire genome is used to optimize only for the primary objective. \textcolor{draft}{After finalizing the search (according to the desired number of epochs) and identifying a potential set of modules for inclusion, we fix the first part of the genome and run a new round of the algorithm that operates exclusively on the second part of the genome.} This second run does not affect the number of modules nor which types of modules are included in the system, rather, it optimizes the load balance according to our secondary objective function by shifting processes around the modules.

\par Same as the genome in our implementation, the final product of this problem is twofold. The first part of the solution is a list of how many modules of each type should be part of a candidate topology that will host our processes. The second is an allocation candidate describing which processes to allocate to each specific module.

\subsection{SP2: MCST Topology Search}
\par Once we know how many processing modules are needed in a candidate topology, we can use this information in conjunction with the graph grammar approach to generate a set of topology graphs representing topologies that satisfy this constraint. We organize possible topology designs in a tree, in which nodes represent partial topologies and directed edges between the nodes represent the application of production rules.

To generate candidate topologies, we start from the initial topology graph (which may be the empty topology) and iteratively apply the MCST technique to explore the tree and evaluate the effects of each rule application. At each iteration, the selection and expansion steps add new nodes of interest to the tree based on what it has learned about each action in the previous iteration. From these nodes of interest, the rollout step simulates the creation of one or more temporary candidate topologies, each of which is evaluated (see SP3 in the next subsection) with respect to the application and topology requirements. The result of each evaluation is back-propagated through the tree and the effects of each rule application are updated. This process is repeated until the maximum number of search epochs (according to a configurable parameter in NetGap) is reached, at which point we stop looking for solutions to SP2, and terminate the search for candidate topologies.

\par At each iteration, the rollout operation explores one branch of the three depth-first by randomly choosing edges that lead to the next level in the tree. Each rollout goes as deep as needed for a terminal configuration, i.e. the topology input to SP3 in this iteration, to be reached. In our context, a terminal configuration is characterized as one that contains the number of processing modules necessary to host the processes (defined in SP1) and that ensures that modules that should communicate can do so.

\par For each node in the tree, we store the visit count (i.e. how many times the node or one of its descendants have been chosen for selection or expansion) and the reward statistics (i.e. the accumulated score of the candidate topologies generated after visiting the node and its descendants) which is what the UCT algorithm \citep{KoSz06} requires as a basis for the selection in each state. In our context, the reward of a partial/final candidate topology is a user-defined function that evaluates the extent to which the topology is able to host the application while satisfying the application and platform requirements (we provide an example of the components of such a reward function in \Cref{sec:evaluation}).
\par  In the current implementation, we have decided on the naive approach for the expansion and rollout task policies, meaning we expand one child per simulation and execute the simulation by choosing actions at random. As a result,  the candidate topology resulting from a rollout task only takes into account possible node additions and interconnection patterns defined by the grammar. Therefore, at this point, candidate topologies do not hold information about how the processes within the nodes communicate or which nodes host which processes. Rather, they are only concerned with the physical connections between the different hardware modules. 

To correctly calculate the merits of a candidate topology with respect to the envisaged application, one needs to map the modules from the solution of SP1 to vertices on the candidate topology graph. We explain this process in the next subsection.

\subsection{SP3: Module Mapping Problem}

Let $T=(V,E,L_V,L_E)$ be a topology graph (see \Cref{def:graph}) generated as a solution for SP2, where $V_{p} \subseteq V$ is the subset of vertices of $T$ labeled as processing modules. Let $M \subset N$  be the set of processing modules included in the solution of SP1 (i.e. $\{n_j \in N | y_i =1\} $ according to \cref{eq:moduleexistence}). Let $w_i: V_p\longrightarrow M$ be a bijective mapping from (labeled) processing vertices in $V_p$ to processing modules in $M$ and $W=\{w_1,...,w_r\}$ be the set of all such possible mappings. Note that for a given topology graph ($V_p$) and a given allocation candidate set $M$, where $|M| = |V_p|= r$ there will be $r!$ (factorial of $r$) possible mappings $w_i$. Solving SP3 is the task of finding the $w_i \in W$ such that when the reward function $f$ is applied to the topology $T$ with the mapping $w_i$, given the requirements expressed by $\mathbb{A}$, the reward is maximized, i.e.:

 \begin{equation}
    \underset{w_i \in W}{argmax} \quad f(T,w_i,\mathbb{A}) 
\end{equation}  

\par As there might exist up to $r!$ possible mappings, finding the optimal solution to the problem is computationally expensive. To approach the problem, we encode it as a genetic algorithm in which the genome of each candidate solution is a list of integers. The index of each position in the list identifies a processing module in $V_{p}$, while the integer at each position identifies a module in $M$. Thus, candidate solution [7,5,3,...] represents that $v_1 \in V_{p}$ is mapped to $m_7 \in M$, $v_2 \in V_{p}$ is mapped to $m_5 \in M$, and so on.
\par At each generation of the genetic algorithm, we evaluate the reward of the current candidate solution in the population according to the desired application and topology requirements. As we have an ordered encoding in which the position of integers in the list matters, we implemented the ordered crossover operator \citep{AbAb12}, and random sequence mutation operators \citep{AbAT12} to generate the offspring.  When SP3 is solved, the reward of the best-performing solution is fed back into the MCST backpropagation stage. 

\section{Evaluation \label{sec:evaluation}}

In this section, we illustrate the scalability of the approach and its flexibility when conceiving multiple platform design alternatives. All measurements have been performed on a laptop equipped with  on a laptop equipped with an Intel Core i7-8750H processor. 

\subsection{Application to an Industrial Use Case}

\par As part of the assessment of the proposed workflow we applied it to a synthetic but realistic use case provided by an industrial partner. In this subsection, we describe the evaluation setup and discuss the results, also showing how the proposed methodology can be used to compare multiple alternative candidate topologies.

\subsubsection{Input description  \label{sec:inputs}} 
\par Below, we describe the use case, its application, and topology requirements as inputs of the topology design process:

\paragraph{Application model description:} The use case consists of an application model constructed to mimic the properties of airborne systems in a concept design stage at our industrial partner. It is divided into two independent parts, a mission-oriented part (MOP) and a flight-critical part (FCP). The mission-oriented part contains 8 periodic processes and 31 periodic messages, and the FCP consists of 91 periodic processes that exchange 629 periodic messages. 

\paragraph{Application requirements:}

\begin{itemize}
    \item \textbf{Security:}  A security breach in the MOP shall not affect the operation of the FCP. Thus, FCP and MOP should not share resources, and processes of each part must run on different network segments. Any cross-segment communication must go through a gateway module.
    \item \textbf{Communication Latency:} The application is time-sensitive, therefore the latency of inter-process messages must be minimized.
    \item \textbf{Resiliency and reconfigurability:} \textcolor{draft}{Communicating modules must have at least two disjoint paths to communicate. Disjoint paths are those that do not share nodes (except for the first and last switches). This setup guarantees that if one switch fails or if the link between two switches breaks, communication will not be completely disrupted.}
    
    \item  \textbf{Resource utilization:} No modules or links should be overloaded, meaning the computation capacity at the processing modules and the communication capacity on network interfaces (therefore on the links) shall not exceed 80\%.
\end{itemize}

\paragraph{Topology grammar and topology requirements:}
\begin{itemize}
    \item \textbf{Module Availability:} For simplicity, we have decided to use homogeneous hardware modules, meaning every hardware module of a given type has the same characteristics. However, in real systems heterogeneous modules  (e.g. with different processing capabilities, number of communication ports, reliability levels, and/or security features) can be mixed and matched depending on the topology description and the graph grammars be adapted accordingly. \Cref{tab:modules} shows the characteristics of each type of candidate hardware module.
    \item \textbf{Topology Grammar:} Due to the application security requirements, the platform configuration is constrained to a segmented mesh topology. In this type of topology, multiple mesh network segments are connected by bridges or gateways.  \Cref{lst:bridgedMeshGrammar} shows the grammar used to describe this configuration. 
    \item \textbf{Cost:} A candidate topology should aim for minimal cost. For simplicity, we have chosen to assign the cost of 10 monetary units (10u) to hardware modules and the cost of 0.1 monetary units (0.1u) to links. This choice reflects the fact that, in common switched networks, adding new links to a system (i.e. physically adding cables and/or configuring the network) tends to be much cheaper than adding new, usually expensive, hardware modules such as switches or processing nodes. These values can be adjusted depending on the type of module and the characteristics of each module (i.e. different modules with different resources and capacities). The maximum cost can also be a constraint, even though this was not done in this example.
    
\end{itemize}

\begin{lstlisting}[caption={A grammar describing a segmented mesh topology with 2 segments} ,mathescape=true,label={lst:bridgedMeshGrammar}, escapechar=\& ,basicstyle=\scriptsize,breaklines=true]
r0: $\phi \Rightarrow G$ &\Comment{\# Adds a gateway node to an empty graph}&
r1: $G_{[0-2]} \Rightarrow$ G $ \leftrightarrow S$ &\Comment{\# Adds a switch S to the graph and connects it to gateway $G$}& &\Comment{ with a maximum of two connections}&
r2: $S_{1[0-14]} \Rightarrow S_1 \leftrightarrow S_2$ &\Comment{\# Adds a switch $S_2$ to the graph and connects it to switch $S$}&
r3: $S_{1[0-14]} , S_{2[0-14]} \Rightarrow S_1 \leftrightarrow S_2$&\Comment{\# Connects two previously unconnected switches $S_1$ and $S_2$ }&
r4: $S_{[0-14]} \Rightarrow S \leftrightarrow M$&\Comment{\# Adds module $M$ to the graph and connects it to switch $S$}&
\end{lstlisting}

\begin{center}
\begin{table}[htb!]
\begin{tabular}{|l|l|l|}
\hline
\multicolumn{1}{|c|}{\textbf{Type}}    & \multicolumn{1}{c|}{\textbf{Resource}} & \multicolumn{1}{c|}{\textbf{Value}} \\
\multicolumn{1}{|c|}{\textbf{of Module}} &  &  \\ \hline
Processing          & Computation capacity                    & 2.7Mops   \\ 
                & Communication capacity                     & 100Mbits/s                          \\
                    & Interface type                    & Full-Duplex                         \\
                    & Cost                              & 10u                                 \\ \hline
Switch              & Communication  capacity                   & 100Mbits/s                          \\
                    & Number of ports                   & 6                                   \\
                    & Interface type                    & Full-Duplex                         \\
                    & Cost                              & 10u                                 \\ \hline
Gateway     & Communication capacity                    & 100Mbits/s                          \\
                    & Number of ports                   & 2                                   \\
                    & Interface type                    & Full-Duplex                         \\
                    & Cost                              & 10u                          \\ \hline
\end{tabular}
\caption{Characteristics of the candidate hardware modules}
\label{tab:modules}
\end{table}
\end{center}
    
\subsubsection{Candidate Topology Evaluation \label{sec:requirements}}
    
    \par Up to this stage, the methodology presented in this paper is general, meaning that any type of topology expressible by the grammar can be generated (an infinite number of such topologies). Therefore, to steer the generation process and make it converge toward a desired solution, one needs to define the topology evaluation function that will be used to calculate the reward of a topology when choosing candidates.
    \textcolor{draft}{ In this use case, topologies are first analyzed for 3 basic structural constraints: Having the right number of processing modules, the presence of 2 disjoint paths between communicating modules, and having 2 network segments. Designs not meeting these criteria are seen as incomplete, receiving the lowest score possible (0 on a scale from 0 to 1).}   
    Topologies that do respect these basic constraints (i.e. have the necessary elements to host application) have their score calculated as the weighted average of the following three terms:
    
    \begin{itemize}
    \item \textbf{Latency score}: The latency score ($l_s$) of a (partial) topology is a value that characterizes how much latency the messages being exchanged by processes mapped to the modules within a given topology are subjected to. In this paper, we represent the latency by a score that combines the mean number of hops in the path of a message($h$), the maximum load in the links ($x_l$), and the number of links that experience a high load ($o$). Mathematically: 
    \begin{equation}
        \label{eq:latency}
        l_s= 2*e^{1-\alpha\: x_l - \beta\: o }/\gamma\: h 
    \end{equation}
        \textcolor{black}{
    where $\alpha$, $\beta$ and $\gamma$ are constants; }
    \begin{equation}
        \textcolor{black}{
        x_l = max\{load_i\; | \; \forall i \in links \},}
    \end{equation}
        \textcolor{black}{
    is the maximum communication load observed in any single link; }
    
    \begin{equation}
        \textcolor{black}{
    o =  |\{i \in links \; | \; load_i > 0.8 \}|}
    \end{equation}
    \textcolor{black}{is the number of overloaded links; and}
     \begin{equation}
     \begin{aligned}
         \textcolor{black}{
         h=mean \{  |\{x \in nodes \; |  x \in shortest\_path (m_{i})  \}|}
        \\   \textcolor{black}{  \forall m_{i} \in messages \}}
     \end{aligned}
     \end{equation}
      
    is the mean number of hops in the path of a message. The negative exponential formulation guarantees convergence towards networks with lower loads and fewer hops. In this paper, we found that $\alpha$, $\beta$, and $\gamma = 1 $ already provide good results, but these weights can be used to trade off latency and load in the requirements.

    \textcolor{draft}{Note that, according to the formulation of \Cref{eq:latency}, the latency score belongs to the interval $[0,1]$, where scores close to $1$ represent lower latencies ("higher is better").}
    
    \item \textbf{Cost:} Different cost functions can be conceived. To begin with, the cost of a topology is just the sum of the costs of its links and nodes, here measured with an ad-hoc unit "u" which can be systematically mapped to monetary cost or energy cost, etc. In this paper, modules (both processing modules and switches) have a cost of 10u and links have a cost of 0.1u.
    
    \item \textbf{Redundant Disjoint Paths:} \textcolor{draft}{The mean number of disjoint paths above two reflects the degree of adherence to the resilience requirements (in this example related to the failure of the links and modules).}
    
    \end{itemize}

\subsubsection{Example of Generated Topologies}

\par
\begin{figure}[htb!]
\centering
\includegraphics[width=0.8\linewidth]{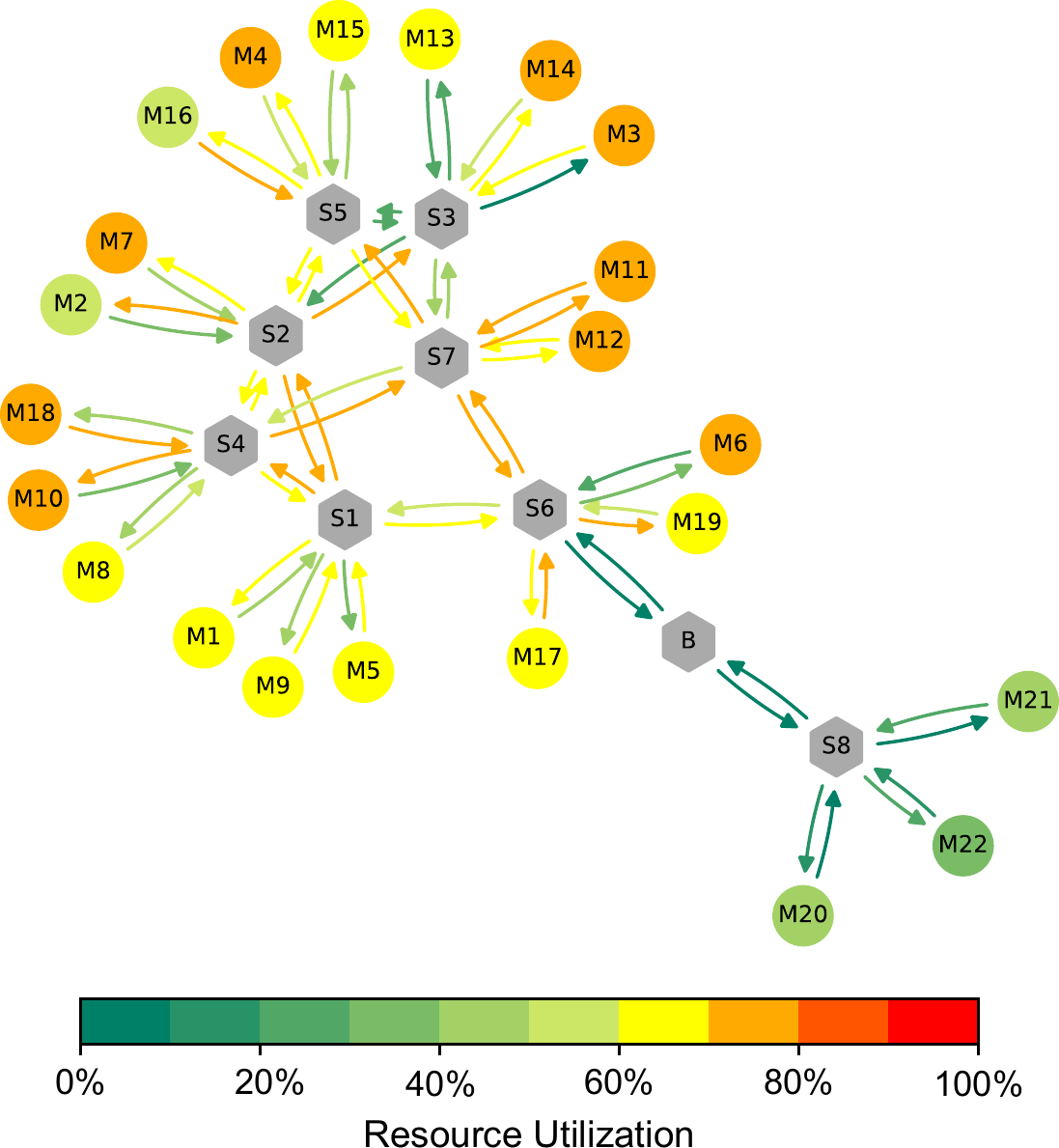}
\caption{Example of a generated candidate topology} \label{fig:ucPlatform} 
\end{figure}

 \Cref{fig:ucPlatform} shows the best candidate topology generated using one iteration of the approach. In the figure, one can easily identify the two network segments connected by the network gateway. One can also see that the lower part of the topology has a complex structure. This shows that the iterative technique we propose for such requirements can generate complex topologies. The total running time to generate this design was 484s.\Cref{tab:NetGAPparameters} shows the tunable parameters of NetGAP.

\begin{table}[htb!]
\centering
\begin{tabular}{|l l|l|}
\hline 
\multicolumn{2}{|l|}{ Parameter} & Value      \\
\hline 
\hline
\multicolumn{3}{|l|}{ Allocation (SP1) GA}  \\
\hline
& Max Generations   & 200 \\
& Max Population Size & 400  \\
& Genome Size (m, q in \cref{sec:allocationProblem}) & (30,91) \\ 
\hline 
\hline
\multicolumn{3}{|l|}{ MCTS (SP2)}  \\
\hline
& Max Search Epochs   & 10000 \\
& Parallel Rollouts   & 1 \\
& UCT exploration constant & 2.8 \\ 
\hline 
\hline
\multicolumn{3}{|l|}{ Mapping (SP3) GA}  \\
\hline
& Max Generations    & 3 \\
& Max Population Size &  50 \\

\hline
\end{tabular}
\caption{NetGAP parameter configuration}
\label{tab:NetGAPparameters}
\end{table}

\par \Cref{tab:ucPlatform} shows the characteristics of the generated topology. As we can see from the table and the figure, both the link and module load constraints are respected, with no link or node above 80\% utilization. Furthermore, all data flows have redundant paths to travel through.

\begin{table}[htb!]
\centering
\begin{tabular}{|l|l|}
\hline
\# of Processing Modules & 22      \\
\# of Switches           & 8      \\
\# of Links              & 70     \\
Cost             & 317u    \\
Network Segments          & 2       \\
Mean Disjoint Paths    & 3.0  \\
Max Node Utilization      & 78.24\% \\
Max Link Utilization      & 79.88\% \\
Mean Hops                 & 3.33     \\ 

\hline
\end{tabular}
\caption{Characteristics of  the example topology in \Cref{fig:ucPlatform}}
\label{tab:ucPlatform}
\end{table}

\subsubsection{Evaluation of Alternative Topologies}

\par
\begin{figure*}[hbt!]
\centering
\includegraphics[width=\textwidth]{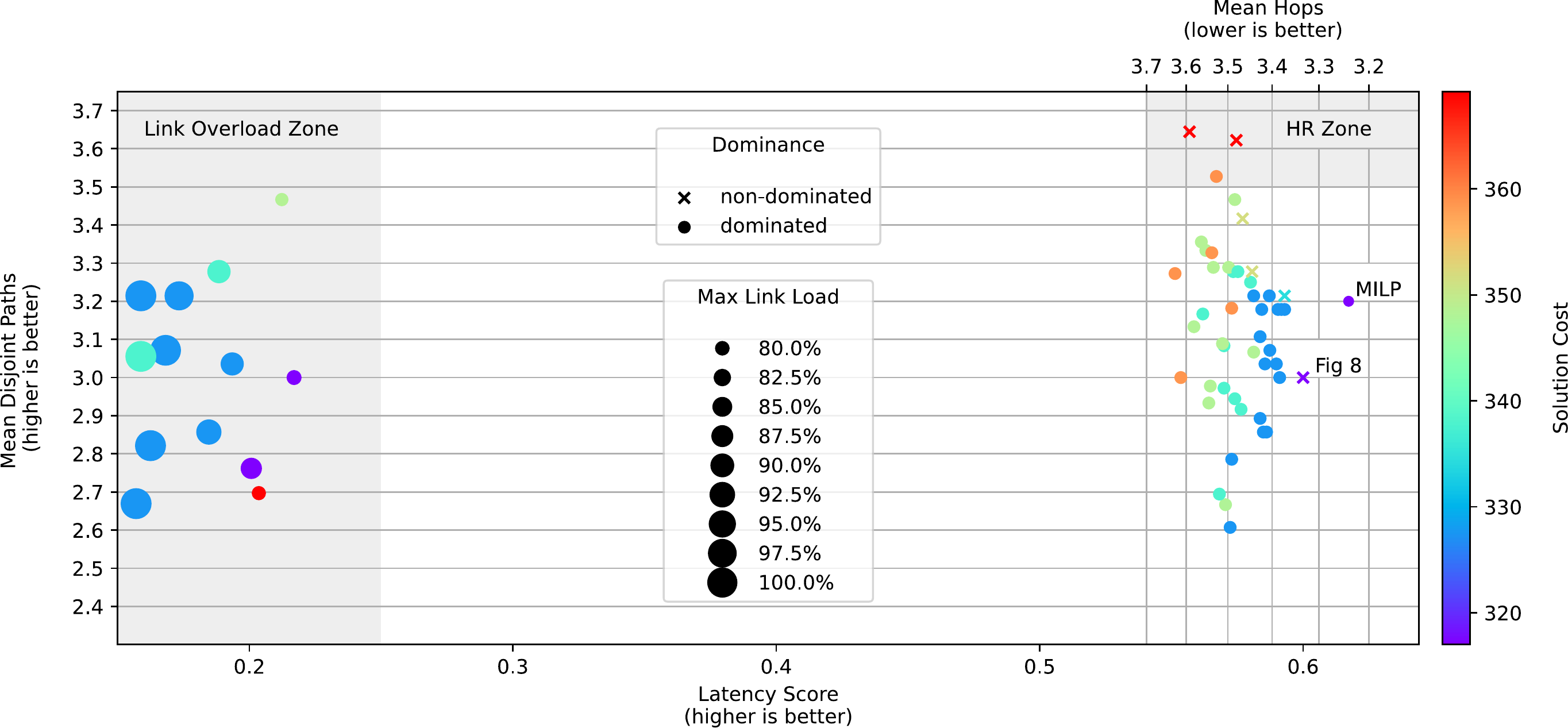}
\caption{Comparison of the characteristics of alternative topologies. Solutions labeled "MILP" correspond to the best solution found through mathematical programming and the one labeled "Fig 8" to the solution in \Cref{fig:ucPlatform}. Non-dominated solutions are plotted as a cross.} 
\label{fig:platformComparsion} 
\end{figure*}

 NetGAP can be used to help concept designers explore the merits and trade-offs of alternative platform configurations. During the concept design, where multiple conflicting requirements and trade-offs are possible, a global view of the merits of different topology choices is a very valuable tool. \Cref{fig:platformComparsion} shows a comparative view of 60 candidate topologies generated using the approach presented in this paper for the same use case. The figure's vertical axis represents the mean number of disjoint paths among communicating modules. The primary (lower) horizontal axis corresponds to the latency score, as per Equation 12. Solutions positioned on the right side showcase a latency score primarily influenced by the average number of hops, which is depicted as a top horizontal axis in the figure. Each data point's size corresponds to the maximum observed load within any link of the topology. Additionally, the color of the data points indicates the cost associated with each topology. The solution labeled "MILP" in the figure corresponds to the best solution found through mathematical programming (discussed in the next subsection) and the one labeled "Fig 8" to the NetGAP solution from Figure 8. Non-dominated solutions are plotted as a cross.

\par  Analyzing the figure, we can identify two broad groups of candidate topologies. The first group, on the left side of the figure (in what we called the link overload zone), is characterized by lower latency scores (under 0.25) that are typically of lower cost and present links that are overutilized (over 80\%). The candidate designs of the second group, to the right of the figure, are characterized by good latency scores, and higher costs on average. As members of this group all have their link utilization below 80\%, the latency score is dominated by the mean number of hops, which we plotted as a second horizontal axis (top axis) so the reader can visualize how the solutions compare in terms of latency.

\par  Among the presented candidates we see topologies that have high resiliency (in what is called the HR zone in the top right). The lower latency score of these candidates is correlated to their higher cost and the presence of more switches and links. 

\par Higher-cost candidates typically exhibit poorer latency scores. This correlation can be attributed to the number of hops in the latency score. Platform configurations that involve more switches generally incur higher costs and introduce additional hops in message pathways.
Similarly, cheaper alternatives tend to have lower redundancy since adding links and switches to the topology increases redundancy but also increases cost. 
Finally, cheaper candidate topologies tend to be the ones that end up more often in the link overload zone since there are fewer switches and links to distribute traffic across, which increases the likelihood of links being overutilized.

\par
\begin{figure}[htb!]
    \centering
    \includegraphics[width=\linewidth]{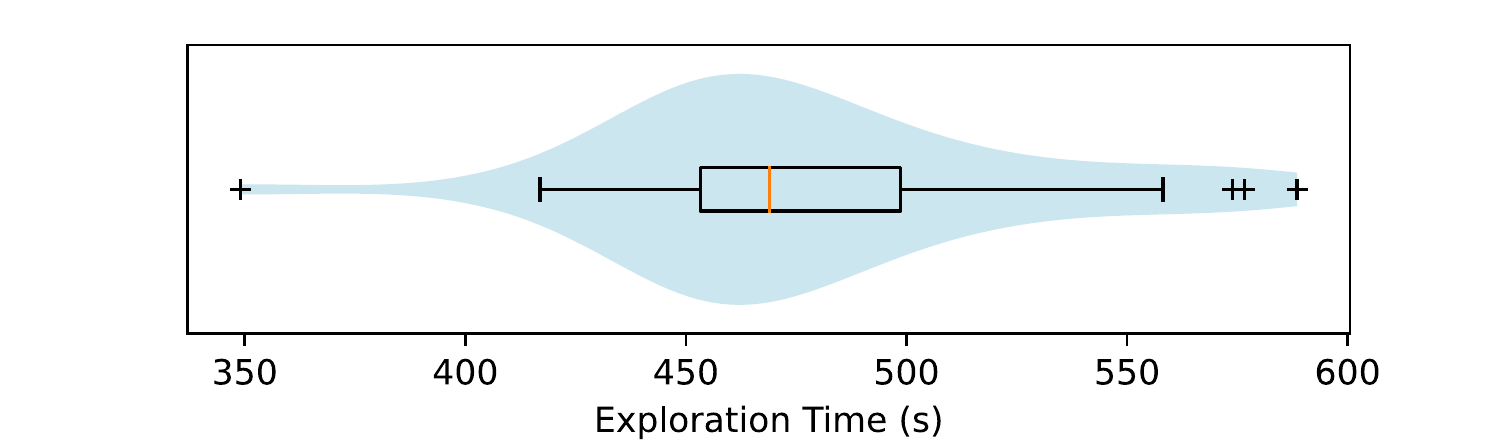}
    \caption{Distribution of the exploration time for the solutions in \Cref{fig:platformComparsion} }
    \label{fig:exploration_time}
\end{figure}

 \textcolor{draft}{\Cref{fig:exploration_time} shows the time distribution of the exploration times for the solutions in \Cref{fig:platformComparsion}. It shows the median time to find a solution (orange bar) and the interquartile ranges. The full experiment (all 60 solutions) took roughly 7 hours, and the mean time to find a solution was 477s.}

\subsubsection{Distance from the Optimal}
\begin{figure}
     \centering
     \begin{subfigure}[b]{0.5\textwidth}
         \centering
         \includegraphics[width=1\textwidth]{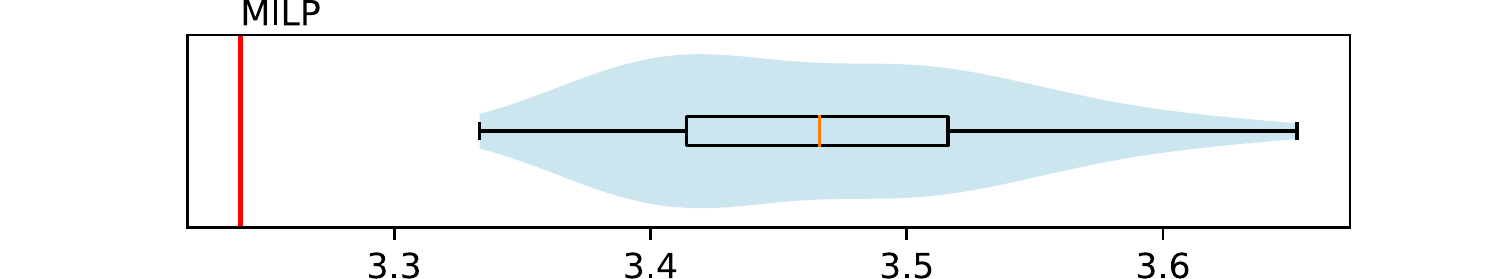}
         \caption{Mean Hops}
     \end{subfigure}
     \hfill

     \begin{subfigure}[b]{0.5\textwidth}
         \centering
         \includegraphics[width=1\textwidth]{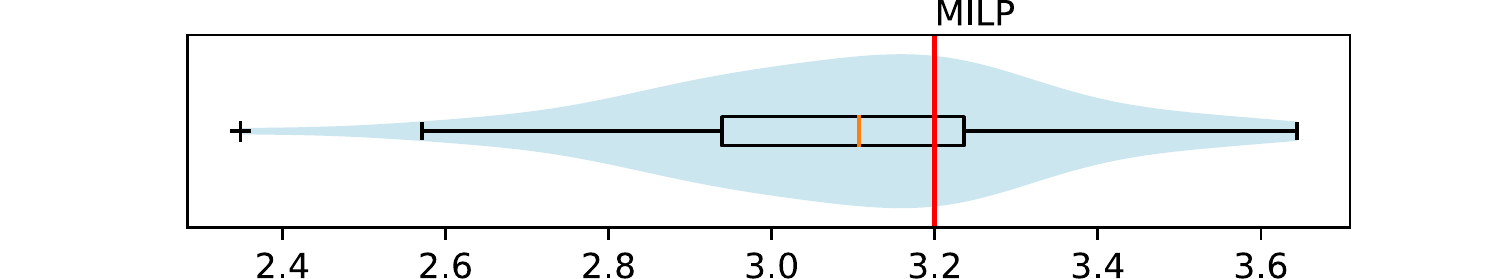}
         \caption{Mean Disjoint Paths}
     \end{subfigure}
 
     \hfil
     \begin{subfigure}[b]{0.5\textwidth}
         \centering
         \includegraphics[width=1\textwidth]{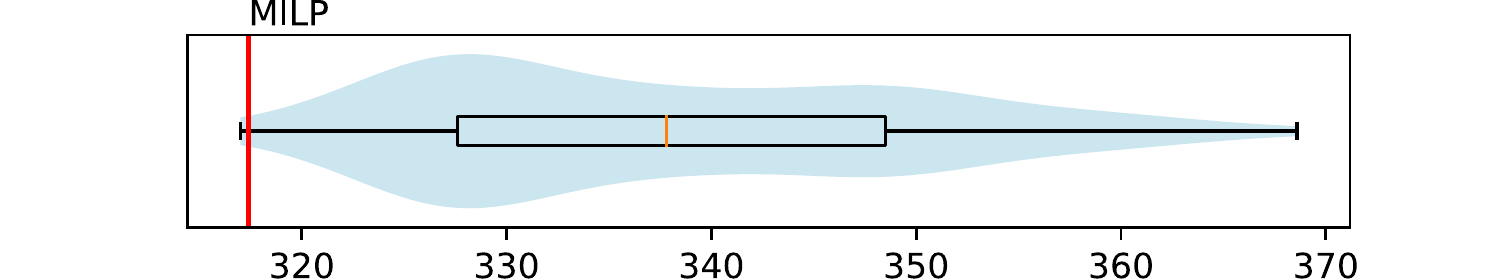}
         \caption{Cost }
     \end{subfigure}
        \caption{Comparison of NetGAP solutions to the MILP solution}
        \label{fig:optimalityComparison}
\end{figure}

\textcolor{draft}{
To study the optimality of our approach, we have modeled the same problem as a MILP and solved it using the Gurobi Solver. After running for over 12 hours on the same machine used for the experiment above, it became evident that the search for the optimal solution would not conclude within a reasonable time frame. At that point we decided to stop the solver, obtaining a solution guaranteed to be within 7.95\% of the LP relaxation optimal. This is the solution labeled MILP in \Cref{fig:platformComparsion}. \Cref{tab:optimalityComparison} shows a comparison between that solution and the one in \Cref{fig:ucPlatform}. \Cref{fig:optimalityComparison} shows  how the feasible solutions of \Cref{fig:platformComparsion} compare to the MILP solution in terms of the used metrics .}

\begin{table}[htb!]
\centering
\begin{tabular}{|l|l|l|}
\hline
Metric & MILP  & NetGAP \\
\hline
\# of Processing Modules & 22 & 22    \\
\# of Switches           & 7  & 7\\
\# of Links              & 74 & 70    \\
Cost                     & 317.4u  & 317.0u  \\
Mean Disjoint Paths        & 3.2 & 3.0  \\
Mean Hops                 & 3.24& 3.33   \\ 
\hline
\end{tabular}
\caption{Comparison between the MILP solution and the solution in \Cref{fig:ucPlatform}}
\label{tab:optimalityComparison}
\end{table}

\textcolor{draft}{
This shows, that NetGAP, on average, approaches the optimal when it comes to the number of mean disjoint paths, but does so by trading off cost and latency. It is important to remember that our aim with NetGAP was never to find optimal solutions but to provide feasible alternatives at the concept stage that could be further improved on and refined at later development stages. }

\subsection{Flexibility of the method \label{sec:flexibility}}

\par \Cref{sec:inputs} described the characteristics of the use-case application model and application/topology requirements used as inputs to NetGAP. During the concept phase, it is very common for requirements to change and for functionality to be added to the platform as the concept development teams evaluate solutions and investigate trade-offs. In this section, we revisit these inputs, and reflect on the flexibility of our methodology to changes to the requirements/applications, to get insight into the amount of work related to each type of requirement change. Most importantly, we discuss whether and under which conditions the results of previous iterations can be re-used to accelerate the search for new topologies when requirements change. This last behavior is especially useful in the last stages of concept development when one (or a few) candidate platforms are satisfactory and only small adjustments are needed. This can also be used to estimate how well the candidate (or even existing) platform can be adapted to reflect changes in the requirements throughout the design process or product lifetime.

\par To start the discussion, we need to analyze the nature of each type of requirement, how each requirement influences the search for topologies, and which stages of the workflow (see \Cref{fig:solutionOverview}) would act upon adjusting for the desired changes. Using this criteria, we can identify two types of requirement changes: those that influence the structure of the topology (e.g. that require the addition of switches and/or links elements to the network), and those that do not. The first type of requirement change leads to more effort in the creation of new candidate topologies.

Where the changed requirement does not change the topology (second type of requirement change), an adjusted solution is obtained much faster.
\par In addition, we can reason about the nature of requirement changes. These can come in two flavors: those that reflect whether or not a requirement should be enforced; and those that relate to the extent to which requirements should be enforced or define thresholds between requirement compliance levels.
\par Next, we discuss each requirement in terms of these two classifications:

\begin{itemize}
    \item \textbf{Cost Requirement:} Changes in the cost requirement will influence how many processing modules (SP1) and how many communication modules and links (SP2) are allowed in topologies. As the solution to SP1  (number of processing modules) is already optimized for cost, we suggest any investigation on changes to the cost requirements to start with SP2 (number of communication modules and links) where more flexibility is possible. This means reusing the current solution to SP1 (process allocation) and allowing the MCTS stage to reflect the change in the old requirements, looking for a less costly topology while keeping the same number of processing modules.
    
    \item \textbf{Latency Requirement:} Similar to cost, latency requirements involve all parts of the problem. While the allocation of processes to processing modules defines the loads on the links adjacent to processing modules (SP1), the presence of switches and the paths between two modules contribute to the number of hops and the load on the other links. For this type of requirement change, we suggest starting directly with SP2 (topology search) and re-using old solutions to SP1 (process allocation) as this can be more flexible and provide more immediate results. 
    
    \item \textbf{Resource Utilization Requirements:} The resource utilization requirement is inherently tied to SP1, therefore, any changes to this requirement (e.g. change the max utilization from 80\% to 90\%), require a new iteration on the whole workflow with limited possibilities to reuse old results (even though some knowledge on the structure of the topology or expected cost can be carried over).
    
    \item \textbf{Segmentation Constraints:} Similar to resource utilization, changes to segmentation requirements, usually for security reasons, imply a total iteration of the workflow and solving all sub-problems anew. Changes to this requirement will direct the process allocation sub-problem to allocate processes from different segments to different modules, and the MCTS topology search to generate structures with different segments separated by gateways. Given the nature of the requirement, no transfer of knowledge or reuse of previous results is possible. 
    
    \item \textbf{Resiliency Requirements:} In our case study, we consider only link redundancy as a measure of resiliency. In this case, changes to the resiliency requirements will only affect the structure of the topologies, potentially requiring new solutions to SP2 to be investigated. If, on the other hand, the use-case considered the process redundancy as important, doing so would require a whole new iteration of the topology search process.
    
    \item \textbf{Changes to the Application:} As the last type of input change, we consider the case of addition or removal of functionalities, or of adaptation of the applications (e.g. adding or removing processes or communication messages). This kind of revision is the one that offers the most possibilities for reuse. Departing from an existing topology, one can first try to map the application into the components where enough resources are available. If such mapping is not possible, one can use an existing topology as an initial point for the NetGAP workflow, allocating the new application processes to new processing modules and adding the necessary links or communication modules to the topology. 
\end{itemize}

\subsection{Scalability of the method}
\label{subsec:scalability}

\textcolor{draft}{
To study the scalability of the solution, we have applied NetGAP to a scaled-up version of the original FCP use case which is 3 times the size of the original use case. This version was constructed to preserve similar characteristics of the original FCP in terms of communication patterns, and message frequencies and sizes. While the original use case contained 91 processes and 629 messages, the new version contains 267 processes and 1887 messages.
For this experiment, we dropped the network segmentation requirement since there is no MOP. The same FCP-only experiment was also run for the original FCP. 
}

\begin{figure}[hbt]
    \centering
    \includegraphics[width=\linewidth]{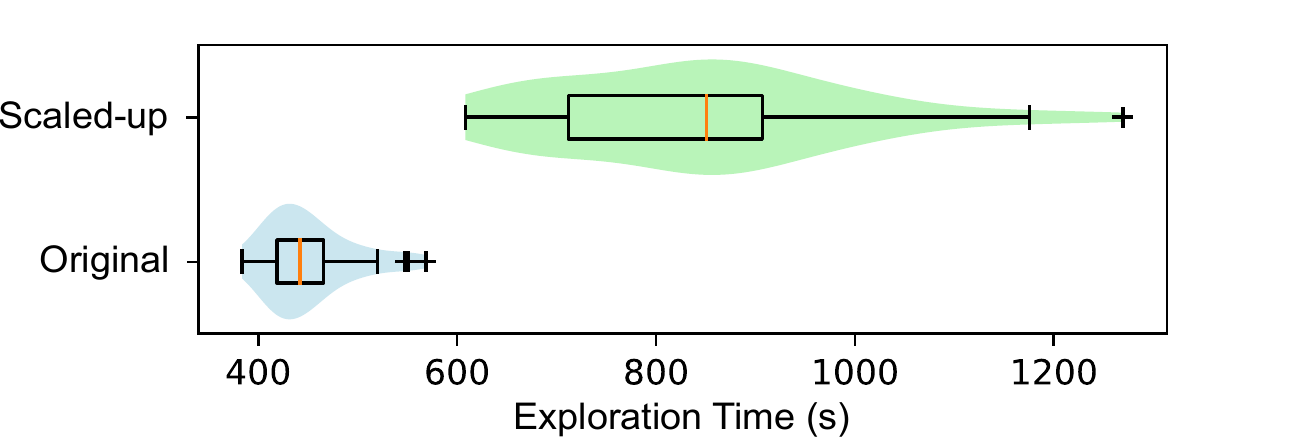}
    \caption{Comparison of exploration time between the scaled-up version of the  FCP use case and the original FCP.}
    \label{fig:time_comparison}
\end{figure}

\textcolor{draft}{\Cref{fig:time_comparison} shows a comparison between the experiments with the two versions.
The analysis shows that the median time to find a feasible solution has roughly doubled (468s to 850s). The same trend can be observed with the interquartile range of the exploration times. This indicates that NetGAP is able to handle much larger sizes than our avionics based use case.}

\par
\begin{figure}[hbt]
    \centering
    \includegraphics[width=\linewidth]{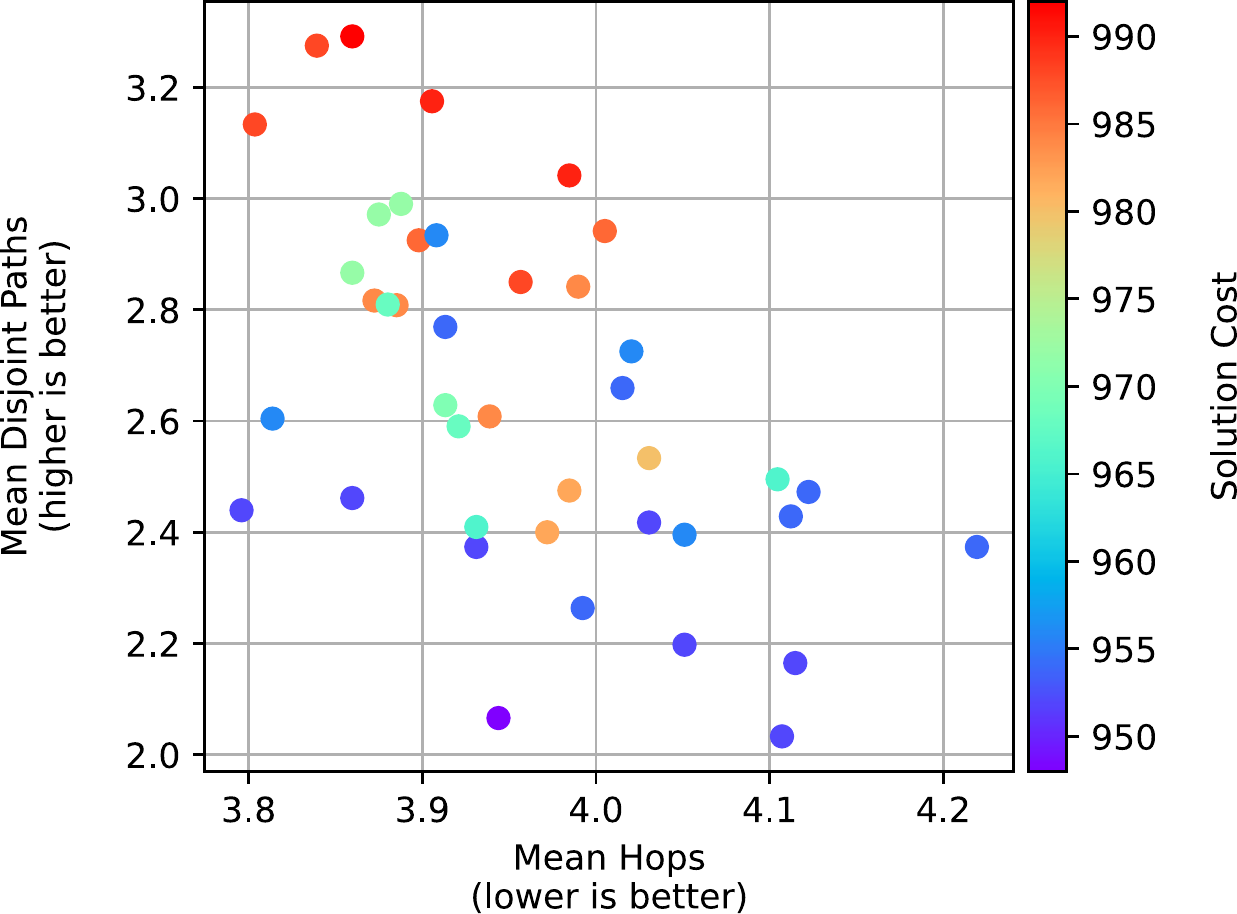}
    \caption{Comparison of the characteristics of alternative topologies for the extended FCP use case. Colors to the cost in monetary units.}
    \label{fig:extended_solutions}
\end{figure}

 \textcolor{draft}{The solutions generated for the scaled-up version experiments are composed of roughly 62 to 64 processing modules and 14 to 16 switches. The trade-off between cost, disjoint paths and mean hops (representing the latency score) for the feasible solutions is shown in \Cref{fig:extended_solutions}.}

\subsection{Further discussions}

In this section, we discuss the merits of our proposed solution in a wider context.

\subsubsection{Industrial Applicability}
\par \textcolor{draft}{The application of our method to an avionics use case was presented and discussed with our industrial partners and was met with appreciation concerning the enhancement of current practices. The industrial experts saw benefit to the concept development of systems-of-systems, where the interface of multiple systems is not always trivial to realize. In this workflow, our approach could be used to identify multiple ways to integrate different sub-system/system modules before moving on to the design of each of these modules. The topology generation approach can also inform the hardware-software co-design process in the next phase of the development.}

\textcolor{draft}{ Since we allow for the definition of case-specific metrics and requirements and use generic techniques like graph grammars our approach should apply to any topology generation problem that can be reduced to a graph manipulation problem. An example is topology generation for the cloud-RAN use case with latency and energy efficiency requirements~\citep{MSTM22} provided all necessary details are shared by the use case owner.}

\textcolor{draft}{To provide some evidence, we have applied the proposed method in a Time Sensitive Network context for message scheduling~\citep{SmNt22}. This technology can appear in multiple domains such as Industry 4.0~\citep{SPFA19} and Automotive~\citep{ZLB+21}.}

\subsubsection{Limitations}

\textcolor{draft}{Any knowledge by the domain engineer helps to create grammars that converge to good solutions. If such knowledge is not available, unrestricted grammars may create large search spaces.
Therefore, it is advisable to opt for fewer but more complex rules instead of many generic rules.}
\par \textcolor{draft}{By decoupling the 3 subproblems, the approach limits the very large size of the search space. While this problem decomposition may lead to overlooking some potential solutions, a one-shot solution to such a problem would not be practical. The inclusion of the assignment problem (SP1) into SP2 and thereby SP3 would have a significant impact on scalability which we could demonstrate in \Cref{subsec:scalability}.}

\section{Conclusions \label{sec:conclusion}}
\par In this paper we have presented NetGAP, a grammar-based method to express and generate candidate platform configurations for networked computer systems,  and the associated three-phase method that helps to analyze the trade-offs. We represent the possible interconnections between the different hardware modules as a set of grammar rules that can be applied in arbitrary sequences to generate different candidate topologies. Using a Monte-Carlo tree search technique, we steer the search toward candidate topologies that observe the requirements of an envisioned application.
\par Through the application of NetGAP to the synthetic version of a realistic use case, we show the ability of the proposed workflow to generate complex candidate topologies and visualize trade-offs. Furthermore, we discuss how this methodology can be used to help concept stage designers compose their own metrics of evaluation, reason about the merits of different topology families, and provide insights on the trade-offs of various candidate designs using the same use case as an example. \textcolor{draft}{Through further examples, we discuss the scalability of the methodology, showing that it scales well with the size of the use case, and conjecture its applicability in domains other than avionics.}
\par The separation of the workflow into three phases allows for modularity, facilitating the reuse of previous solutions when dealing with changes in the requirements (as discussed in section \Cref{sec:flexibility}). However, dealing with requirement changes is not a straightforward process and requires knowledge about the impact of each requirement in each phase of the workflow. For example, if the latency requirements are made more precise, shifting from average latencies to maximum latencies, one needs to adopt further analyses on a given candidate topology using earlier known techniques (e.g. \cite{SmBN21}).
\par Since the grammar can be created to provide arbitrarily large and complex system architectures the approach does not limit the engineers in terms of expressivity. The limitation would, however, be in the process and time taken to converge to "good enough" topologies. Using the non-trivial application we chose, we ascertain that the method is fast enough for deployment at the concept stage, a fact that was confirmed when presenting the approach to our industrial partners in the aeronautics domain. 
\par The work in this paper could be extended with the adoption of different and more complex heuristics for the rollout task policies within the MCST framework. The choice of intelligent rules to guide the rollout phase of the simulation can significantly increase the performance of MCST methods by steering the exploration towards better solutions, hence improving the quality of the explored partial designs. As an example, one can steer the rollout towards adding more switches and links to the partial designs when redundancy or load balancing is a priority at that stage. Similarly, the rollout could be steered towards the addition of certain types of processing modules in case the mere number of modules is deemed insufficient at that stage. Although the size and composition of hardware modules dictate a large part of energy consumption, there is room for including more specific metrics for energy efficiency trade-offs. These and other approaches to rank and allocate priorities to different teams' requirements within an organization appear possible to build upon the basic method that we have devised and illustrated in this paper.
\par  \textcolor{draft}{Finally, using Multi-Objective Evolutionary Algorithms (MOEAs) to solve SP3 could be a promising direction for future research. MOEAs are designed to handle problems with multiple conflicting objectives, which could be beneficial for SP3 given the competing optimization objectives it considers.}

\section*{Acknowledgements \label{sec:ack}}
    This work was supported by Sweden's Innovation Agency --  Vinnova, as part of the national projects on aeronautics, NFFP7, project CLASSICS (NFFP7-04890).  The authors wish to thank the industrial partners from Saab AB for their valuable technical input and the use case used in the paper.

\bibliographystyle{elsarticle-harv}
\bibliography{sections/mybib}

\end{document}